\documentclass[12pt,pra,aps,amssymb,amsfonts,amsmath,tightenlines]{revtex4}
\usepackage{dsfont}
\usepackage{graphicx}
\usepackage{amssymb,amsfonts,amsthm}
\usepackage{color}
\usepackage{verbatim}
\usepackage[normalem]{ulem}
\usepackage{enumerate}
\usepackage{mathrsfs}
\usepackage{bm}

\newtheorem{Proposition}{Proposition} 
\newtheorem{Lemma}{Lemma}

\newtheorem{Theorem}{Theorem}
\newcommand{\half}{\mbox{$\textstyle \frac{1}{2}$}}

\newcommand{\re}{\mbox{$\rm e$}}

\newcommand{\rd}{\mbox{$\rm d$}}

\newcommand{\norm}[1]{\left\lVert#1\right\rVert}

\begin{document}

\title{L\'evy-Ito Models in Finance}

\author{ George Bouzianis$^1$, Lane~P.~Hughston$^1$, \\ Sebastian Jaimungal$^2$ and Leandro S\'anchez-Betancourt${}^3$}

\affiliation{
$^1$Department of Computing, Goldsmiths University of London,\\ New Cross, London SE14\,6NW, United Kingdom \\
$^2$Department of Statistical Sciences, University of Toronto, Toronto M5S 3G3, Canada \\
$^3$Mathematical Institute, University of Oxford,  Oxford OX2 6GG, United Kingdom
}

\begin{abstract}
\noindent We present an overview of the broad class of financial models in which the prices of assets are L\'evy-Ito processes driven by an $n$-dimensional Brownian motion and an independent Poisson random measure. The Poisson random measure is associated with an $n$-dimensional L\'evy process. Each model consists of a pricing kernel, a money market account, and one or more risky assets.  We show how the excess rate of return above the interest rate can be calculated for risky assets in such models, thus showing the relationship between risk and return when asset prices have jumps. The framework is applied to a variety of asset classes, allowing one to construct new models as well as interesting generalizations of familiar models. 
\vspace{-0.2cm}
\\
\begin{center}
{\scriptsize {\bf Key words: Asset pricing, interest rate models, foreign exchange, risk premium, risk aversion, \\pricing kernels, 
L\'evy processes,  L\'evy measure,  Poisson random measure,  Siegel's paradox, Vasicek model.
} }
\end{center}
\end{abstract}

\maketitle
\section{Introduction}
\label{sec:LIP}
\noindent  Pricing models driven by L\'evy processes have been considered by many authors  
 \cite{Andersen Lipton, Brody Hughston Mackie 2012, Chan, Cont Tankov, Eberlein 1995, Eberlein 1998, Eberlein 2001, Gerber, Hubalek, Kuchler Tappe, Madan Seneta 1990, Schoutens}. We are concerned  here  with a  broader family of pricing models, namely, the so-called L\'evy-Ito models. Such models are driven both by a Brownian motion and a Poisson random measure, where the Poisson random measure is associated with an underlying L\'evy process. The L\'evy-Ito class is general enough to include many familiar models as special cases, such as models driven by L\'evy processes, yet offers ample opportunity for the creation of new models, while retaining substantial tractability. 

The need for a systematic theory of L\'evy-Ito models in finance is plain, for if an asset price is driven by a L\'evy process, then the price process of an option or other derivative based on that asset cannot itself in general be represented by a L\'evy model, but it can typically be represented 
by a L\'evy-Ito model, provided that the payoff is reasonably well behaved; and as we know well  \cite{Black Scholes, Merton 1974}, most securities and other financial assets, both corporate and sovereign, can be viewed as complex options based on the cash flows associated with one or more simpler underlying assets. 
Our intention in the material that follows is to present the theory of L\'evy-Ito models for asset pricing from a unified point of view, working  in the real-world measure and emphasizing the role of the excess rate of return. We  present a number of specific examples of tractable L\'evy-Ito models, all suitable for implementation, ranging across a variety of asset classes,  including equities, interest rates, and foreign exchange, illustrating the flexibility of the modelling framework. 

The structure of the article is as follows. In Section II we present a synopsis of the L\'evy-Ito calculus. The approach that we adopt is rigorous, but we try to avoid abstractions and material unnecessary for financial applications. 
In Section III, we present a theory of risky assets driven by L\'evy-Ito processes. We assume the existence of a L\'evy-Ito pricing kernel (state price density) of the form \eqref{Levy Ito pricing kernel RA}, then in Proposition 1 we deduce the general form that the price process of a risky asset takes in a L\'evy-Ito market model. We comment on the nature of the excess rate of return above the short rate of interest in a L\'evy-Ito setting, and in equation \eqref{excess rate of return above the interest rate} we show that the excess rate of return per unit of jump intensity can be expressed as the product of a random volatility and a random market-price-of-risk for each admissible jump vector of the L\'evy process associated with the Poisson random measure. 

In Section IV we develop a theory of interest rate models in a L\'evy-Ito framework, and in Propositions 2 and 3 we work out expressions for the money market account and the discount bond system in such a model. The resulting theory is more general than the well-known interest rate models driven by Brownian motion and pure-jump L\'evy processes, yet remains tractable and suitable for implementation. As an example of a L\'evy-Ito interest rate model, in Section V we present an extension of the Vasicek model to the L\'evy-Ito category, summarized in Proposition 4, generalizing results of \cite{Vasicek, Cairns, Norberg 2004, Brody Hughston Meier 2018, Eberlein Kallsen} and others.  

In Section VI we show that the so-called ``chaotic'' interest rate models \cite{Hughston Rafailidis 2005, Brody Hughston 2004, Rafailidis} lift naturally to the  L\'evy-Ito category. In particular, in Proposition 5, we show that the pricing kernel in a L\'evy-Ito model for interest rates can be written as the conditional variance of a random variable that admits a Wiener chaos expansion. Then we work out explicit formulae for the discount bond prices in a class of second-order chaos models. In Section VII we specialize to the case of so-called factorizable chaos models, and in Proposition 6 we show how such models can be calibrated to an arbitrary initial interest-rate term structure. 

Finally, in  Section VIII we consider L\'evy-Ito models for foreign exchange, and in Proposition 7 we present a general expression for the exchange rate matrix for any number of currencies in a L\'evy-Ito setting. We conclude with an analysis of the excess rates of returns that can be exhibited by reciprocal exchange rate pairs in a multi-currency situation. We prove that in an $N$-currency geometric Brownian model driven by $n$ Brownian motions ($N \geq 2, \, n \geq 2$) one can choose the market-price-of-risk vectors in such a way that the excess rate of return above the interest-rate differential is strictly positive for each currency pair. In the case of a two-currency model driven by a single Brownian motion this result is known as Siegel's paradox \cite{Black 1990, Siegel}, and here we have shown that the Siegel condition can be satisfied for each exchange rate pair in a multi-currency Brownian market model. We are also able to present examples of multi-currency L\'evy models in which the Siegel condition is satisfied. This leads us to conjecture that the Siegel condition can be satisfied in any multi-currency L\'evy-Ito model with an appropriate choice of the risk-aversion functions. 

\section{L\'evy-Ito Calculus}

\noindent We begin with an overview of the L\'evy-Ito calculus, which acts as the work horse of the theory, and give examples of typical calculations. Although many of the ideas that follow are well known, it is not easy to locate any one treatment elsewhere in the literature that offers a concise synopsis of the main results of the  theory required for applications to financial modelling. As a consequence this material may be of interest in its own right along with the applications discussed in later sections of the article.
 
In L\'evy-Ito models, the prices of financial assets are driven collectively by an $n$-dimensional Brownian motion together with a 
Poisson random measure defined on 
$ \mathbb R_0^n \times \mathbb R^+ $.  Here we write  $\mathbb R_0^n =  \mathbb R^n - \{0\}$ and $\mathbb R^+ $ denotes as the non-negative real numbers. If $A$ and $B$ are subsets of a set $E$, then we set $A - B = A \cap B^c$ where $B^c = \{\omega \in E : \omega \notin B\}$. We refer to $\mathbb R_0^n$ as the state space of the Poisson random measure.  In the class of models with which we work, the Poisson random measure is associated with a  pre-specified  $n$-dimensional L\'evy process. That is, we assume the existence of an underlying L\'evy process of dimension $n$, and we consider the Poisson random measure determined by this process. We emphasize that the class of models driven by the Poisson random measure associated with a L\'evy process is much larger than the class of models driven by the L\'evy process itself. 
For simplicity, we first discuss the situation where the Brownian motion and the Poisson random measure are each of dimension one; the higher dimensional case can then be reconstructed by analogy with a slight adjustment of notation. When we model the price dynamics of a single risky asset, we find that for some purposes a one-dimensional model will suffice; but when we consider collections of assets, as one must for interest rates and foreign exchange, then the need for L\'evy-Ito models with higher-dimensional state spaces becomes apparent. 

We assume that the reader has some familiarity with the mathematical theory of 
L\'evy processes, as represented in works such as  
\cite{Applebaum, Bertoin, Boyarchenko Levendorskii 2002, Cinlar, Eberlein Kallsen, Ikeda Watanabe, Jeanblanc et al, Kijima 2002, Kyprianou, Oksendal, Protter, Sato, Tankov}, and the applications of L\'evy processes in finance theory. 
We fix a probability space $(\Omega, \mathscr{F} , \mathbb{P})$ and let $\{\xi_t\}_{t\geq0}$ be a 
one-dimensional L\'evy process. In our notation for stochastic processes, curly brackets $\{ \cdot \}$ signify an indexed set of random variables. The index space is usually indicated explicitly when the process is defined, but can be dropped later for brevity, unless we wish to draw attention to the index set. Thus, we can now refer to the process $\{\xi_t\}$, since we have already mentioned the index set $\{t \in \mathbb R^+\}$. The same convention applies to filtrations. It is often taken as part of the definition of L\'evy process that the process has the c\`adl\`ag property; that is to say, there exists a set  $\Omega_1 \in \mathscr F$ with $\mathbb P(\Omega_1) = 1$ on which the sample paths of $\{\xi_t\}$ are right-continuous with left limits.  
 
Let us write $\mathscr{B}(\mathbb{R})$  for the Borel sigma algebra generated by the open sets of 
$\mathbb{R}$. $\mathscr{B}(\mathbb{R}^+)$, $\mathscr{B}(\mathbb{R}_0)$ and $\mathscr{B}(\mathbb{R}^+_0)$ are defined similarly.  It is well known that a one-dimensional L\'evy process $\{\xi_t\}$ admits a so-called L\'evy-Ito decomposition (\cite{Sato}, theorem 19.2) of the form 
\begin{eqnarray} 
\xi_t =   \alpha t \,  + \beta \, W_t + \int_{0}^{t}\int_{|x| \in (0,1)} x \, \tilde{{N}}({\rm d}x, {\rm d}s) + \int_{0}^{t}\int_{|x| \geq 1} x \, {N}({\rm d}x, {\rm d}s)\,.
\label{Levy Ito decomposition}
\end{eqnarray}
Here $\alpha$ and $\beta$ are constants, $\{W_t\}_{t\geq0}$ is a standard Brownian motion, and $\{{N}({\rm d}x, {\rm d}t)\}$ is an independent Poisson random measure. More specifically,  $\{{N}({\rm d}x, {\rm d}t)\}$ is the random measure on $ \mathbb R_0 \times \mathbb R^+ \times \Omega$ defined for  $A \, \in \, \mathscr{B}(\mathbb{R}_0)$, ${t\geq0}$ and $\omega \in \Omega_1$ (with $\Omega_1$ as above) by
\begin{eqnarray} 
{ N}(A, [0,t], \omega) = \#\{ s \in [0,t] : \Delta \xi_s(\omega) \in A\},
\label{Levy measure definition}
\end{eqnarray}
and we set ${ N}(A, [0,t], \omega) = 0$ for $\omega \notin \Omega_1$. Thus, for any outcome of chance $\omega \in \Omega_1$, the value of the random variable ${ N}(A, [0,t])$ measures the number of jumps occurring in the time interval $[0, t]$ for which the jump size lies in the set $A$. 
The so-called L\'evy measure $\nu(A)$ associated with $\{{N}({\rm d}x, {\rm d}t)\}$ is then defined for $A \in \mathscr{B}(\mathbb{R}_0)$ by
\begin{eqnarray} 
\mathbb{E} \left[ { N}(A, [0,t]) \right] = \nu(A)\,t\, .
\end{eqnarray}
By a L\'evy measure on $\mathbb R$ we mean a sigma-finite measure $\{\nu({\rm d}x)\}$ on $(\mathbb R, \mathscr{B(\mathbb{R})})$, not necessarily finite, such that 
$\nu(\{0\}) = 0$ and
\begin{eqnarray} 
\int_{x} \min \,(1, |x|^2) \, \nu({\rm d}x )< \infty\,,
\end{eqnarray}
where the integral is understood to be taken over $\mathbb R$. 

For any set $A \in \mathscr{B}(\mathbb{R})$ let us write $\bar A$ for the closure of $A$.  Then we say that $A \in \mathscr{B}(\mathbb{R}_0)$ is bounded below if $0 \notin \bar A$. For example, if $0 < a < b$ then $(a, b)$ and $[a, b)$ are bounded below, as are $(-b, -a)$ and $(-b, -a]$. We note that if $A$ is bounded below, then $N(A, [0,t]) < \infty$ almost surely for all $t\geq 0$, and $\nu(A) < \infty$.  The compensated Poisson random measure $\tilde N(A, [0,t])$ for such a set is then defined by 
\begin{eqnarray} 
\tilde N(A, [0,t]) =  N(A, [0,t]) -  \nu(A) \, t .
\label{compensated measure} 
\end{eqnarray}
If $A$ is bounded below then if we let $t$ vary we obtain a Poisson process $\{N(A, [0,t])\}_{t\geq 0}$ with rate $\nu(A)$; then $\{N(A, [0,t]) - \nu(A) t \}_{t\geq 0}$ is the corresponding compensated Poisson process. In the case of the compensated measure it  is customary shorthand to write 
\begin{eqnarray} 
\tilde{ N}({\rm d}x, {\rm d}s) = {N}({\rm d}x, {\rm d}s) - \nu({\rm d}x)\,{\rm d}s\,,
\label{compensated PRM}
\end{eqnarray}
but it should be noted that the integral with respect to $\tilde{ N}({\rm d}x, {\rm d}s)$ in the third term on the right side of  (\ref{Levy Ito decomposition}) cannot in general be split into separate terms by use of (\ref{compensated PRM}).  Rather, the term as a whole is defined by a limiting procedure (\cite{Sato}, page 120). In particular, given any decreasing sequence $\{\epsilon_n\}_{n \in \mathbb N}$ such that $\epsilon_n \in \mathbb R^+_0$,  $\epsilon_1 < 1$, and $\epsilon_n \downarrow0$, the sets $\{A_n\}_{n \in \mathbb N}$ defined by $A_n = \{ x : \epsilon_n < |x| < 1\}$ are each bounded below. Then for each $t>0$ we set
\begin{eqnarray} 
\int_{0}^{t}\int_{|x| \in (0,1)} x \, \tilde{{N}}({\rm d}x, {\rm d}s)
= \lim_{n \to \infty} \int_{0}^{t}\int_{A_n} x \, \tilde{{N}}({\rm d}x, {\rm d}s)\, .
\label{compensated integral}
\end{eqnarray}
If for any given value of $n$ we use \eqref{compensated PRM} to work out the integral on the right hand side of 
\eqref{compensated integral}, the result is a square-integrable random variable $Y_n \in L^2(\Omega, \mathscr F, \mathbb P)$. Then we can show that $\{Y_n\}$ is a Cauchy sequence that converges  to an element $Y \in L^2(\Omega, \mathscr F, \mathbb P)$, and this is the definition of the integral on the left-hand side of equation \eqref{compensated integral}. 

From the foregoing we see that if we specify a L\'evy process on $(\Omega, \mathscr{F} , \mathbb{P})$, we determine a Brownian motion $\{W_t\}$ and a Poisson random measure  $\{{N}({\rm d}x, {\rm d}t)\}$ with L\'evy measure $\{\nu({\rm d}x)\}$. 
By a \textit{L\'evy-Ito process} on $(\Omega, \mathscr{F} , \mathbb{P})$ we mean a process $\{Y_t\}$ of the form
\begin{eqnarray} 
Y_t = Y_0 + \int_{0}^{t} \alpha_s \, {\rm d}s + \int_{0}^{t} \beta_s \, {\rm d}W_s + \int_{0}^{t}\int_{|x|\in (0,1)} \gamma_s(x) \, \tilde{{N}}({\rm d}x, {\rm d}s) + \int_{0}^{t}\int_{|x| \geq 1} \delta_s(x) \, {N}({\rm d}x, {\rm d}s). \nonumber \\
\label{Levy Ito process general form}
\end{eqnarray}
The integrands appearing in the various terms here have to satisfy certain conditions to ensure that the relevant integrals are well defined. More specifically, we require that $\{ \alpha_t \}_{t \geq 0}$, $\{ \beta_t \}_{t \geq 0}$, $\{ \gamma_t(x)\}_{t \geq 0,\,|x| \in [0,1)}$ and $\{ \delta_t(x)\}_{t \geq 0,\,|x| \in [1,\infty)}$ are predictable and that the following holds for all $t \geq 0$: 
\begin{eqnarray} 
 \mathbb{P} \left[  \,
\int_{0}^{t} \left(  |\alpha_s| +  \beta_s^{\,2} +
 \int_{|x|<1} \gamma^{\,2}_s(x) \, \nu({\rm d}x)  \right) \, {\rm d}s < \infty \right ] = 1\,.
 \label{P condition}
\end{eqnarray}

Here we recall that a process $\{ \phi_t \}_{t \geq 0}$ on a probability space  $(\Omega, \mathscr{F} , \mathbb{P})$ with filtration $\{ \mathscr{F}_t \}_{t \geq 0}$ is said to be predictable if the map $\phi: \mathbb R^{+} \times \Omega \rightarrow \mathbb R$ is measurable with respect to the so-called predictable $\sigma$-algebra $\mathscr P$, which is the $\sigma$-algebra generated by all left-continuous adapted processes on $(\Omega, \mathscr{F} , \mathbb{P})$. More precisely, $\mathscr P$
is the $\sigma$-algebra generated by all maps of the form $\theta:  \mathbb R^{+} \times \Omega \rightarrow \mathbb R$ such that (a)  for fixed $\omega \in \Omega$ the map $t \mapsto \theta_t(\omega)$ is left-continuous, and ($b$) for fixed $t \in \mathbb R^{+}$ the map $\omega \mapsto \theta_t(\omega)$ is $ \mathscr{F}_{t^-}$ \!measurable, where we define
\begin{eqnarray}
\mathscr{F}_{t^-} = \sigma \left( \bigcup_{\,0 \leq s < t} \! \mathscr{F}_{s} \right) .
\end{eqnarray}

For map-valued processes, such as the processes $\{ \gamma_t(x)\}$ and $\{ \delta_t(x)\}$ appearing in the final two terms of (\ref{Levy Ito process general form}), we need a somewhat more general definition.  
Let $A \in \mathscr{B}(\mathbb R)$ be the domain in $\mathbb R$ on which the maps are to be defined, which could be $\mathbb R$ itself. The predictable $\sigma$-algebra $\mathscr{P}_A$ is then defined to be the $\sigma$-algebra generated by all maps of the form $\theta: A \times \mathbb R^{+} \times \Omega \rightarrow \mathbb R$ such that  ($a$) for fixed $t \in \mathbb R^{+}$ the map $(x,\omega) \mapsto \theta_t(x,\omega)$ is $\mathscr{B}(A) \otimes \mathscr{F}_{t^-}$ \!measurable, and ($b$) for fixed $x \in A$ and $\omega \in \Omega$ the map $t \mapsto \theta_t(x,\omega)$ is left-continuous.

Any process 
$\{\theta_t(x)\}_{t\geq 0, \,x \in A}$ defined by a $\mathscr{P}_A$-measurable map $\theta: A \times \mathbb R^{+} \times \Omega \rightarrow \mathbb R$ is said to be predictable. If $\theta$ is predictable, then the process $t \mapsto \theta_t(x)$ is adapted for each $x \in A$. 

For $A \in \mathscr {B} (\mathbb R)$  we define $\mathscr{P}_2(A, \mathbb R^+)$ to be the set of all mappings (modulo equivalence) of the form 
$\theta: A \times \mathbb R^{+} \times \Omega \rightarrow \mathbb R$  such that  
$\{\theta_t(x)\}$ is predictable and the condition 
\begin{eqnarray} 
\mathbb{P} \left[  \int_{0}^{t} \int_{A} \theta^{\,2}_s(x) \, \nu({\rm d}x) \, {\rm d}s < \infty \right ] = 1
\label{P_2 condition}
\end{eqnarray}
holds for $t \geq 0$.  Two such processes are taken to be equivalent if they coincide almost surely with respect to $ \nu \times {\rm Leb} \times \mathbb P$ on $\mathbb{R}\otimes \mathbb{R}^+ \otimes \Omega$. Thus, if $\theta$ and $\theta'$ are two such maps, we say that they are equivalent if the set
$\{ ( x, t, \omega)  \in  \mathbb{R}\otimes \mathbb{R}^+ \otimes \Omega   \, : \, \theta (x, t, \omega) \neq \theta' (x, t, \omega) \}$
is of measure zero with respect to $ \nu \times {\rm Leb} \times \mathbb P$.
We note, in particular, by virtue of \eqref{P condition}, that the process $\{\gamma_t(x)\}$ appearing in equation  \eqref{Levy Ito process general form} is in $\mathscr{P}_2(A, \mathbb R^{+})$ for $A = \{x \in \mathbb R : |x| < 1\}$.

In calculations, one often finds it convenient to write (\ref{Levy Ito process general form}) in differential form. Then the initial condition is implicit and we have
\begin{eqnarray} 
{\rm d}Y_t =  \alpha_t \, {\rm d}t +  \beta_t \, {\rm d}W_t + \int_{|x|\in (0,1)} \gamma_t(x) \, \tilde{{N}}({\rm d}x, {\rm d}t) + \int_{|x|\geq1} \delta_t(x) \, {N}({\rm d}x, {\rm d}t)\,.
\label{Levy Ito process differential form}
\end{eqnarray}
As in the classical Ito calculus, the meaning of such a differential form comes from the corresponding integral expression.  
 
We proceed to consider a generalized version of Ito's lemma applicable to  L\'evy-Ito processes. First we recall the form that Ito's lemma takes for a one-dimensional semimartingale on a probability space $(\Omega, \mathscr{F} , \mathbb{P})$ with filtration $\{ \mathscr{F}_t \}_{t \geq 0}$ (Protter \cite{Protter}, theorem 32).  
\vspace{3mm}
\begin{Theorem} 
Let $\{Y_t\}_{t \geq 0}$ be a semimartigale. Let the map $F: \mathbb R \rightarrow \mathbb R$ admit a continuous second derivative and write $F{'}(x)$ and  $F{''}(x)$ for the first and second derivatives of $F$ at $x \in \mathbb R$.
Then
\begin{align} 
F(Y_t) & = F(Y_0) + \int_{0}^{t}  F{'}(Y_{s^{-}}) \,{\rm d}Y_s  
+ \tfrac{1}{2}\, \int_{0}^{t} F{''}(Y_{s^{-}})\,\,{\rm d}[\,Y,Y]^c_s \nonumber
\\ & \hspace{1cm}+ \sum_{0<s \leq t} \left\{F(Y_{s} ) - F(Y_{s^{-}}) - \Delta Y_s \, F'(Y_{s^{-}})\right\}.
\label{Ito's formula for semimartigales}
\end{align}
\end{Theorem} 
\vspace{3mm}
\noindent Here, for any process $\{X_t\}_{t \geq 0}$ admitting left limits we set 
\begin{eqnarray}
X_{t^-} =  \lim_{\,s\, \uparrow \,t}X_s \, \mathds{1}(t>0)+ X_{0} \, \mathds{1}(t=0),
\end{eqnarray}
and we write $\Delta X_t = X_t - X_{t^-}$ for $t \geq 0$.  The time integrals are taken over the interval $(0, t]$. We use the notation $\{[Y,Y]_t\}_{t\geq 0}$ to denote the quadratic variation process, defined by 
\begin{eqnarray}
[Y,Y]_t = Y^2_t - 2  \int_{0}^{t} Y_{s^{-}} \,{\rm d}Y_s \,.
\end{eqnarray}
Since the quadratic variation is nondecreasing and has right continuous paths such that $\Delta [\,Y,Y]_t =  (\Delta Y_t)^2$, it can decomposed into a continuous part and a discontinuous part, and we write $\{[\,Y,Y]^c_t\}$ for the continuous part. 

Now let $\{Y_t\}_{t \geq 0}$ be a L\'evy-Ito process given in the form (\ref{Levy Ito process general form}). Some simplification can then be achieved for the expression for $\{F(Y_t)\}_{t \geq 0}$. First, we can separate the continuous terms and the jump terms, and use familiar formulae from the continuous version of Ito's lemma.  Then we can re-express the infinite sum over the jumps of the process as an integral with respect to the Poisson random measure, splitting the contributions from the small jumps and the large jumps. The result is 
\begin{align} 
F(Y_t) & = F(Y_0) + \int_{0}^{t} \left[ \alpha_s F{'}(Y_{s^{-}}) + \tfrac{1}{2}\, \beta_s^2 F{''}(Y_{s^{-}})\right] {\rm d}s + \int_{0}^{t} \beta_s F{'}(Y_{s^{-}})\,{\rm d}W_s   \nonumber
\\&   
+ \int_{0}^{t}\int_{|x|\in (0,1)} \gamma_s(x) F{'}(Y_{s^{-}}) \, \tilde{{N}}({\rm d}x, {\rm d}s)  \nonumber
\\&   
+ \int_{0}^{t} \int_{|x|\in (0,1)}\left[F(Y_{s^{-}} + \gamma_s(x)) - F(Y_{s^{-}}) - \gamma_s(x) \, F'(Y_{s^{-}}) \right] {N}({\rm d}x, {\rm d}s ) \nonumber
\\ &  + \int_{0}^{t} \int_{|x| \geq 1}\left[F(Y_{s^{-}} + \delta_s(x)) - F(Y_{s^{-}}) \right] {N}({\rm d}x, {\rm d}s )
\label{Ito's formula before moderation}
\end{align}
The version of Ito's formula given by \eqref{Ito's formula before moderation} is valid for any L\'evy-Ito process, and for any continuously twice differentiable function $F$. 

For financial modelling we need a further assumption. This concerns the infinite sum over the small jumps implicit in the penultimate term of  \eqref{Ito's formula before moderation}. The point is that for financial applications (and other applications as well) it is often the case that we need to introduce constructions involving local martingales.  
This means that we need to ensure that the integral with respect to the Poisson random measure over the small jumps in the next to last term in \eqref{Ito's formula before moderation} can be replaced with an integral with respect to the compensated Poisson random measure. In short, we would like to have necessary and sufficient conditions for $\{ F(Y_{t})\}$ itself to be a 
L\'evy-Ito process. 

First we observe that the integrand $\{\gamma_s(x) \, F'(Y_{s^{-}})\}$ belongs to $\mathscr P_2\{(-1,1), \mathbb R^+ \}$. This ensures that the integral in the antepenultimate term of \eqref{Ito's formula before moderation} is almost surely finite. Indeed, if we multiply any process $\{\eta_t(x)\} \in \mathscr P_2\{(-1,1), \mathbb R^+ \}$ with a left-continuous adapted process $\{X_t\}$ then the result $\{X_t \, \eta_t(x)\}$ also lies in $\mathscr P_2\{(-1,1), \mathbb R^+ \}$. 
Thus to ensure that the Poisson random measure in the penultimate term can be compensated, it is necessary and sufficient to assume that $F$ and  $\{Y_t\}$ taken together are such that  
\begin{eqnarray}
\{F(Y_{s^{-}} + \gamma_s(x)) - F(Y_{s^{-}})\} \in \mathscr P_2\{(-1,1), \mathbb R^+ \}.
\label{assumption}
\end{eqnarray}
\noindent The effect of this condition is to moderate the impact of the small jumps in such a way that the integral of $F(Y_{s^{-}} + \gamma_s(x)) - F(Y_{s^{-}})$ is well defined with respect to the compensated Poisson random measure. 
For \eqref{assumption} to hold it suffices that either (a) $F$ is bounded, or (b) $\{\gamma_t(x)\}$ is locally bounded in the sense that for all $t>0$ it holds that
\begin{eqnarray} 
\mathbb{P} \left[ \sup_{0 \leq s \leq t} \, \sup_{0 \leq |x |< 1} \left | \gamma_s(x) \right | < \infty \right] = 1.
\label{sup sup}
\end{eqnarray}
From a modelling perspective assumption \eqref{sup sup} is not unreasonably restrictive, for it simply means that the jumps in the process $\{F(Y_t)\}$ are not unduly sensitive to small jumps in the underlying L\'evy process. 
With condition \eqref{assumption} in hand, a further simplification results in the form of Ito's lemma, which then takes the following form.
\vspace{.3cm}
\begin{Theorem} Let $\{Y_t\}_{t\geq 0}$ be a L\'evy-Ito process of the form \eqref{Levy Ito process general form}, let the map $F: \mathbb R \to \mathbb R$ admit a continuous second derivative, and assume that \eqref{assumption} holds. Then 
\begin{align} 
F(Y_t) & = F(Y_0) + \int_{0}^{t} \left[ \alpha_s F{'}(Y_{s^{-}}) + \tfrac{1}{2}\, \beta_s^2 F{''}(Y_{s^{-}})\right] {\rm d}s + \int_{0}^{t} \beta_s F{'}(Y_{s^{-}})\,{\rm d}W_s \nonumber
\\ & \hspace{1cm}+ \int_{0}^{t} \int_{|x|<1}\left[F(Y_{s^{-}} + \gamma_s(x)) - F(Y_{s^{-}}) - \gamma_s(x)F{'}(Y_{s^{-}})\right]\, \nu({\rm d}x) \, {\rm d}s \nonumber
\\ & \hspace{1cm}+ \int_{0}^{t} \int_{|x|\in (0,1)}\left[F(Y_{s^{-}} + \gamma_s(x)) - F(Y_{s^{-}})\right]\, \tilde{N}({\rm d}x, {\rm d}s ) \nonumber
\\ & \hspace{1cm}+ \int_{0}^{t} \int_{|x| \geq 1}\left[F(Y_{s^{-}} + \delta_s(x)) - F(Y_{s^{-}})\right]\, N({\rm d}x, {\rm d}s )\,.
\label{Ito's formula}
\end{align}
\end{Theorem}
\vspace{.3cm}
\noindent This is the version of Ito's lemma proved in Applebaum \cite{Applebaum}, theorem 4.4.7, and also (in a somewhat more general form) in Ikeda \& Watanabe \cite{Ikeda Watanabe}, theorem 5.1.  Going forward, we shall make use of Ito's lemma in this form, making the necessary assumptions without further comment. 
We observe that if $\{Y_t\}$ is a L\'evy-Ito process and if $F$ is continuously twice-differentiable, then we find that $\{F(Y_t)\}$ is a  
L\'evy-Ito process. In more detail, we have
\begin{align}
F(Y_t) - F(Y_0) & = \int_{0}^{t} A_s \, {\rm d}s + \int_{0}^{t} B_s  \, 
{\rm d}W_s + \int_{0}^{t} \int_{|x| \in (0,1)}C_s(x)\,\tilde{N}({\rm d}x, {\rm d}s ) \nonumber
\\ & \hspace{1cm} + \int_{0}^{t} \int_{|x| \geq 1} D_s(x) \, N({\rm d}x, {\rm d}s ),
\label{ABCD}
\end{align} 
where we define
\begin{align}
A_s = &  \,\alpha_s \, F{'}(Y_{s^{-}}) + \frac{1}{2} \beta_s^{\,2} \, F{''}(Y_{s^{-}}) \nonumber\\ &+ \int_{|x| < 1} \left[F(Y_{s^{-}} + \gamma_s(x)) - F(Y_{s^{-}}) - \gamma_s(x)F{'}(Y_{s^{-}})\right] \, \nu ( {\rm d}x)\,, \nonumber
\end{align}
\begin{eqnarray}
B_s = \, \beta_s \, F{'}(Y_{s^{-}})\,, \quad C_s(x)= \, F(Y_{s^{-}} + \gamma_s(x)) - F(Y_{s^{-}})\,, \nonumber
\end{eqnarray}
\begin{eqnarray}
D_s(x)= \, F(Y_{s^{-}} + \delta_s(x)) - F(Y_{s^{-}})\,.
\end{eqnarray}
In particular, it follows as a consequence of  Theorem 2 that for all $t\geq 0$ it holds that
\begin{eqnarray}
 \mathbb{P} \left[ \int_{0}^{t} \left( |A_s|  + B_s^{\,2}  +  \int_{|x|<1} C^{\,2}_s(x) \, \nu({\rm d}x)\, \right){\rm d}s < \infty\right] = 1\,.
\end{eqnarray}
 Note that in calculations it can be useful to write  (\ref{Ito's formula}) in differential form, and we have 
\begin{align} 
{\rm d} F(Y_t) & =  \left[ \alpha_t \, F{'}(Y_{t^{-}}) + \tfrac{1}{2} \beta_t^{\,2} \, F{''}(Y_{t^{-}})\right] {\rm d}t + \beta_t \,F{'}(Y_{t^{-}}) \,{\rm d}W_t \nonumber
\\ & \hspace{1cm} + \int_{|x|<1}\left[F(Y_{t^{-}} + \gamma_t(x)) - F(Y_{t^{-}}) - \gamma_t(x)F{'}(Y_{t^{-}})\right]\, \nu({\rm d}x) {\rm d}t \nonumber
\\ &\hspace{1cm} + \int_{|x|\in (0,1)}\left[F(Y_{t^{-}} + \gamma_t(x)) - F(Y_{t^{-}})\right]\, \tilde{N}({\rm d}x, {\rm d}t ) \nonumber
\\ & \hspace{1cm} + \int_{|x| \geq 1}\left[F(Y_{t^{-}} + \delta_t(x)) - F(Y_{t^{-}})\right]\, N({\rm d}x, {\rm d}t ) \, .
\end{align}
\vspace{0.1cm}

\noindent \textbf{Example 1}. As a step towards the construction of a pricing model we consider the problem of solving a stochastic differential equation of the form
\begin{eqnarray}
{\rm d} Z_{t} = Z_{t^{-}} \left[ \, \mu_t \, {\rm d}t + \int_{|x|\in (0,1)} \Gamma_t(x)\tilde{N}({\rm d}x, {\rm d}t ) + \int_{|x| \geq 1}\Delta_t(x)N({\rm d}x, {\rm d}t ) \right ], 
\label{Levy Ito SDE example}
\end{eqnarray}
given the processes $\{ \mu_t \}_{t \geq 0}$, $\{\Gamma_t(x)\}_{t \geq 0,\,|x| \in (0,1)}$ and 
$\{\Delta_t(x)\}_{t \geq 0,\,|x| \in [1, \infty)}$ as inputs, along with a strictly positive  initial value $Z_0$.  
We assume that $\{ \mu_t \}$ is predictable and such that 
\begin{eqnarray} 
\mathbb{P} \left[ \int_0^t \left | \mu_s \right | {\rm d}s < \infty \right ] = 1
\end{eqnarray}
for $t \geq 0$, $\{\Gamma_t(x)\} \in \mathscr{P}_2\{(-1,1), \mathbb R^+\}$, and $\{\Delta_t(x)\}$ is predictable. We also assume  that 
\begin{eqnarray} 
\mathbb{P} \left[ \sup_{0 \leq s \leq t} \, \sup_{0 \leq| x |< 1}  \Gamma_s(x) < \infty \right] = 1
\label{Gamma sup sup}
\end{eqnarray}
and 
\begin{eqnarray} 
\mathbb{P} \left[ \inf_{0 \leq s \leq t} \, \inf_{0 \leq|x| < 1}  \Gamma_s(x) > -1 \right] = 1, \quad  
\mathbb{P} \left[ \inf_{0 \leq s \leq t} \, \inf_{1 \leq |x |< \infty}  \Delta_s(x) > -1 \right] = 1
\label{inf inf}
\end{eqnarray}
for $t \geq 0$. The latter two conditions ensure that $\{Z_t\}$ will not jump to a negative value or to a value that is arbitrarily close to zero; whereas 
\eqref{Gamma sup sup} and the first part of \eqref{inf inf} ensure that $\{\Gamma_t(x)\}$ is locally bounded, so we can apply Ito's formula to obtain 
\begin{align}
{\rm d} \log Z_t  & =  \,    \mu_t \, {\rm d}t + \int_{|x|<1} \big( \log\left(1+\Gamma_t(x)\right) - \Gamma_t(x) \big) \, \nu ({\rm d}x){\rm d}t \nonumber
\\ & + \int_{|x|\in (0,1)}  \log\left(1+\Gamma_t(x)\right)   \,  \tilde{N}({\rm d}x, {\rm d}t )  + \int_{|x| \geq 1}   \log\left(1+\Delta_t(x)\right)   \,  N({\rm d}x, {\rm d}t )\, .
\end{align}
Then the solution of (\ref{Levy Ito SDE example}) is given by 
\begin{align}
& Z_t  = Z_0  \exp \bigg \{ \int_{0}^{t} \mu_s \, {\rm d}s  
+ \int_{0}^{t} \int_{|x|<1} \big( \log(1+\Gamma_s(x)) - \Gamma_s(x) \big) \, \nu ({\rm d}x) \, {\rm d}s \bigg\}  
\nonumber \\
& \times \exp \bigg \{ \int_{0}^{t}  \int_{|x|\in (0,1)} \log(1+\Gamma_s(x)) \, \tilde{N}({\rm d}x, {\rm d}s ) + \int_{0}^{t}  \int_{|x| \geq 1} \log(1+\Delta_s(x)) \, N({\rm d}x, {\rm d}s )\bigg\} .
\label{Example 1 solution}
\end{align}
We remark, finally, that in applications, it is often convenient to write \eqref{Example 1 solution} in the alternative but equivalent form 
\begin{align}
Z_t = Z_0 & \exp \bigg \{  \int_{0}^{t} \mu_s \, {\rm d}s  
- \int_{0}^{t}  \int_{|x|<1}   \left( \re^{\gamma_s(x)}-1-\gamma_s(x) \right) \, \nu ({\rm d}x) {\rm d}s \bigg\}  \nonumber \\ 
& \times \exp \bigg \{ \int_{0}^{t}  \int_{|x| \in (0,1)} \gamma_s(x) \tilde{N}({\rm d}x, {\rm d}s ) + \int_{0}^{t}   \int_{|x| \geq 1} \delta_s(x) N({\rm d}x, {\rm d}s )\bigg\} , 
\end{align}
where $\gamma_t(x) = \log\left(1+\Gamma_t(x)\right)$ and $\delta_t (x) = \log\left(1+\Delta_t(x)\right)$.

\vspace{0.5cm}

\noindent \textbf {Example 2}. Next we consider the construction of exponential martingales in a L\'evy-Ito framework. For this purpose,  instead of (\ref{Levy Ito SDE example}) we look at the slightly modified equation
\begin{eqnarray} 
{\rm d} Z_{t} = Z_{t^{-}} \left[ \int_{|x| \in (0,1)} \Gamma_t(x)\tilde{N}({\rm d}x, {\rm d}t ) + \int_{|x| \geq 1}\Delta_t(x) \tilde{N}({\rm d}x, {\rm d}t ) \right],
\label{Levy Ito SDE modified equation}
\end{eqnarray}
the difference being that there is no drift term and we use the compensated Poisson random measure in both integrals. This opens up the possibility that we can make $\{Z_t\}$ a local martingale and even a martingale. The treatment of the $|x| \in (0,1)$ integral is just as in the previous example. In order for the compensator term to be defined in the 
$|x| \geq 1$ integral 
we require  that
\begin{eqnarray}
\mathbb P \left[ \int_{0}^{t} \int_{|x| \geq 1} \left | {\Delta_s(x)} \right |\,  \, \nu ({\rm d}x) \, {\rm d}s <\infty \right] = 1 \, ,
\label{condition on Delta}
\end{eqnarray}
for all $t \geq 0$. As a consequence of \eqref{condition on Delta} we can then write \eqref{Levy Ito SDE modified equation} in the form
\begin{align} 
{\rm d} Z_{t} = Z_{t^{-}} \left[ -\int_{|x| \geq 1} {\Delta_t(x)} \,  \nu ({\rm d}x)  {\rm d}t + \int_{|x| \in (0,1)} \Gamma_t(x)\tilde{N}({\rm d}x, {\rm d}t ) + \int_{|x| \geq 1}\Delta_t(x) N({\rm d}x, {\rm d}t ) \right].
\label{Levy Ito SDE  further modified equation}
\end{align}
But we see that \eqref{Levy Ito SDE further modified equation} is of the form \eqref{Levy Ito SDE example}, with 
\begin{eqnarray} 
\mu_t = - \int_{|x| \geq 1} {\Delta_t(x)}  \, \nu ({\rm d}x)\, .
\end{eqnarray}
It follows by equation \eqref{Example 1 solution} in Example 1 that the solution takes the form
\begin{align}
Z_t &= \,Z_0  \exp \bigg \{ \int_{0}^{t}  \int_{|x| \in (0,1)} \log(1+\Gamma_s(x)) \, \tilde{N}({\rm d}x, {\rm d}s ) 
+ \int_{0}^{t} \int_{|x|<1} \big( \log(1+\Gamma_s(x)) - \Gamma_s(x) \big) \, \nu ({\rm d}x) \, {\rm d}s \bigg\}  
\nonumber \\
& \times \exp \bigg \{  \int_{0}^{t}  \int_{|x| \geq 1} \log(1+\Delta_s(x)) \, N({\rm d}x, {\rm d}s ) - \int_{0}^{t} \int_{|x| \geq 1} {\Delta_s(x)}  \, \nu ({\rm d}x) \, {\rm d}s )\bigg\} .
\end{align}
 Next, we observe that if the process $\{\Delta_t(x)\}$ also satisfies
\begin{eqnarray}
\mathbb P \left[ \int_{0}^{t} \int_{|x| \geq 1} \left |\,{\log (1 + \Delta_s(x)})\,\right |  \, \nu ({\rm d}x) \, {\rm d}s <\infty \right] = 1 
\label{further condition on Delta}
\end{eqnarray}
for $t\geq 0$, then one can introduce a compensator into the stochastic integral for $|x| \geq 1$ as well, and the expression for  $\{Z_t\}$ can be put into the symmetrical form
\begin{align}
Z_t &= \, Z_0  \exp \bigg \{ \int_{0}^{t}  \int_{|x| \in (0,1)} \log(1+\Gamma_s(x)) \, \tilde{N}({\rm d}x, {\rm d}s ) 
+ \int_{0}^{t} \int_{|x|<1} \big( \log(1+\Gamma_s(x)) - \Gamma_s(x) \big) \, \nu ({\rm d}x) \, {\rm d}s \bigg\}  
\nonumber \\
& \times \exp \bigg \{ \int_{0}^{t}  \int_{|x|\geq1} \log(1+\Delta_s(x)) \, \tilde{N}({\rm d}x, {\rm d}s ) 
+ \int_{0}^{t} \int_{|x|\geq1} \big( \log(1+\Delta_s(x)) - \Delta_s(x) \big) \, \nu ({\rm d}x) \, {\rm d}s \bigg\}  .
\label{local martingale}
\end{align}
But \eqref{further condition on Delta} is satisfied under the conditions that we have imposed, since
\begin{align}
\left |\,{\log (1 + \Delta_s(x)})\,\right | 
\leq   \left |\,\log \left(1 + \inf_{0 \leq s \leq t} \, \inf_{1 \leq |x |< \infty}  \Delta_s(x)\right) \,\right | \mathds{1}( \Delta_s(x) \leq 0)
+ \Delta_s(x) \mathds{1}( \Delta_s(x) > 0), 
\end{align}
which follows by \eqref{inf inf} and the inequality $\log (1 + x) \leq x$ for $x > -1$. Then we can write (\ref{local martingale}) in the form
\begin{align}
Z_t &= \, Z_0 \, \exp \bigg \{ \int_{0}^{t}  \int_{|x|>0} \log(1+\Sigma_s(x)) \, \tilde{N}({\rm d}x, {\rm d}s ) 
+ \int_{0}^{t} \int_{x} \big( \log(1+\Sigma_s(x)) - \Sigma_s(x) \big) \, \nu ({\rm d}x) \, {\rm d}s \bigg\} , 
\label{symmetrical form}
\end{align}
where 
\begin{eqnarray}
\Sigma_t(x) = \mathds{1}(|x|\in [0,1)) \, \Gamma_t(x)  +  \mathds{1}(|x| \geq 1) \, \Delta_t(x),
\end{eqnarray}
and  \eqref{Levy Ito SDE modified equation} takes the compact form
\begin{eqnarray}
{\rm d} Z_t  = Z_{t^{-}}   \int_{|x|>0}   \Sigma_t(x)\tilde{N}({\rm d}x, {\rm d}t ) \, . 
\label{SDE for exp mart}
\end{eqnarray}
It follows that $\{Z_t\}$ is a local martingale, and since $\{Z_t\}$ is strictly positive a sufficient condition to ensure that it is a martingale is that 
\begin{eqnarray}
\mathbb{E} [Z_t] = Z_0
\end{eqnarray}
for all $t \geq 0$.
The solution \eqref{symmetrical form} can also be written as
\begin{align}
Z_t & = Z_0 \,  \exp \bigg \{ \int_{0}^{t}  \int_{|x| >0}  \sigma_s(x) \tilde{N}({\rm d}x, {\rm d}s ) 
- \int_{0}^{t}  \int_{x}   \left( \re^{\sigma_s(x)}-1-\sigma_s(x) \right) \, \nu ({\rm d}x) \, {\rm d}s \bigg\}, 
\label{alternative symmetrical form}
\end{align}
where 
\begin{eqnarray}
\sigma_t(x) = \log (1 + \Sigma_t(x) ),
\end{eqnarray}
and thus
\begin{eqnarray}
{\rm d} Z_t  = Z_{t^{-}}   \int_{|x|>0}  \left( \re^{\sigma_s(x)}-1 \right) \tilde{N}({\rm d}x, {\rm d}t ) \, . 
\label{SDE for exp mart}
\end{eqnarray}

We observe in the case of a pure-jump L\'evy-Ito process, the volatility appears in two distinct forms. We call $\{\sigma_t(x)\}_{t \geq 0}$ the exponential volatility and  $\{\Sigma_t(x)\}_{t \geq 0}$ the dynamical volatility.  The condition \eqref{condition on Delta} that we have imposed on the dynamic volatility translates into an analogous condition on the exponential volatility, namely
\begin{eqnarray} 
\mathbb{P} \left[ \int_{0}^{t} \int_{|x|\geq 1} \left | \re^{\sigma_s (x)} - 1 \right | \, \nu({\rm d}x) \, {\rm d}s < \infty \right] = 1\,. 
\label{first condition on sigma}
\end{eqnarray}
Since $\nu([1, \infty)) < \infty$, this condition can be simplified by use of the identity 
\begin{eqnarray}
 \left | \re^{\sigma_s (x)} - 1 \right | =  \mathds{1}(\sigma_s (x)>0) (\re^{\sigma_s (x)} -1) + \mathds{1}(\sigma_s (x)<0) (1 - \re^{\sigma_s (x)})\, ,
\end{eqnarray}
and thus in place of \eqref{first condition on sigma} we can write
\begin{eqnarray} 
\mathbb{P} \left[ \int_{0}^{t} \int_{|x|\geq 1}  \re^{\sigma_s (x)} \, \nu({\rm d}x) \, {\rm d}s < \infty \right] = 1 \, . 
\label{alternative condition on sigma}
\end{eqnarray}
Thus we require the L\'evy measure to satisfy a type of exponential moment condition. Such conditions routinely arise in L\'evy models for asset pricing, so it is not surprising to see similar conditions arising in the context of 
L\'evy-Ito models.  Note that by \eqref{further condition on Delta} we also have
\begin{eqnarray}
\mathbb P \left[ \int_{0}^{t} \int_{|x| \geq 1} \left |\,\sigma_s (x) \,\right |  \, \nu ({\rm d}x) \, {\rm d}s <\infty \right] = 1\, .
\end{eqnarray}
%

\section{Risky assets}
\label{sec:RA}
\noindent
We begin with a few general remarks about asset pricing, following which we introduce a class of L\'evy-Ito models for risky assets. Let $({\Omega},{\mathscr F},{\mathbb P})$ be a probability space equipped with a filtration 
$\{{\mathscr F_t}\}_{t\geq 0} $  satisfying the usual conditions, where ${\mathbb P}$ is the real-world measure. We write ${\mathbb E}_t [\,\cdot\,]$ for conditional expectation with respect to 
$\mathscr{F}_t$ under ${\mathbb P}$.  Equalities and inequalities between
random variables are generally understood to hold ${\mathbb P}$-almost-surely. We write $\rm m\mathscr F_t $ 
for the space of  $\mathscr F_t $-measurable ${\mathds R}$-valued  random variables. 
 Price processes are 
modelled by semimartingales, denominated in units of a fixed base currency. In particular, we assume that price processes are c\`adl\`ag.

We assume the existence of 
a so-called pricing kernel, following the definition of \cite{Brody Hughston 2018}, by which we mean a semimartingale $\{\pi_t\}_{t\geq0}$, satisfying (i) $\pi_t >0$ for $t\geq0$, (ii) ${\mathbb E}\, [\,\pi_t\,] < \infty$ for
$t\geq0$, and (iii) $\liminf_{t\to\infty} 
{\mathbb E}\,[\pi_t]=0$. The pricing kernel has the defining property that  if an asset with price $\{S_t\}_{t\geq0}$ delivers a single random nonnegative cash flow $X_T \in \rm m\mathscr F_T $ at time $T$ such that $ \pi_T X_T$ is integrable, and if the asset derives its value entirely from that cash flow, 
then its value at any time $t\geq 0$ is given by 
\begin{eqnarray}
S_{t} =  {\mathds 1}_{\{ t< T\} } \frac{1}{\pi_t}\, {\mathbb E}_t [\pi_T X_T ] \, .
\label{pricing formula}
\end{eqnarray}
Note that the value of the asset drops to zero when the cash flow occurs and remains at that value thereafter. It is known from Jobert \& Rogers \cite{JR} that if a pricing operator is linear and satisfies some reasonable consistency conditions that impose a mild form of absence of arbitrage, then it is of the form \eqref{pricing formula}. More generally, in the case of an asset that delivers  a stream of nonnegative dividends we introduce a random measure $\Delta({\rm d}t)$ on $\mathbb R^+$ with the property that for any 
$A \in \mathscr B (\mathbb R^+)$ the total dividend paid over the time period $A$ is given by  $\Delta(A)$. We require  that $\Delta(A) < \infty$ almost surely for any bounded set $A$.  Then for any asset with value process $(S_t)_{t\geq0}$ and dividend $\Delta({\rm d}t)$, the pricing kernel must be such that the deflated total value process $({S}^*_t)_{t\geq0}$, defined by
\begin{eqnarray} 
{S}^*_t = \pi_t S_t + \int_0^t \pi_s \,\Delta(\rd s)\, ,
\label{deflated total value}	
\end{eqnarray}
is a martingale. Then if the so-called transversality condition 
\begin{eqnarray} 
\liminf_{t \to \infty}\, [\pi_t S_t ] = 0
\label{transversality}	
\end{eqnarray}
is satisfied, we say that the asset derives its value entirely from the dividend stream, and a calculation shows that
\begin{eqnarray} 
{S}_t =  \frac{1}{\pi_t}\, \mathbb E_t \left [\int_t^{\infty}\! \pi_s \,\Delta(\rd s) \right ] .
\label{value from dividends}		
\end{eqnarray}
If an asset derives its value from a cash flow $X_T$ at time $T$ we have
$\Delta(\rd t) = X_T\,\delta_T(\rd t)$, where $\delta_T(\rd t)$ denotes the Dirac measure concentrated with unit mass at time $T$. It is then
a straightforward exercise using the martingale property of the total deflated value  \eqref{deflated total value} to show that \eqref{value from dividends} reduces to \eqref{pricing formula} in that case.

For many applications we find it convenient to assume the existence of a unit-initialized money-market asset 
with value  $\{B_t\}_{t\geq0}$ such that 
\begin{eqnarray} 
B_t = \exp \left[\int_0^t r_s \, \rd s\right] , 
\label{money market account formula}
\end{eqnarray}
for some specified short rate  $\{r_t\}_{t\geq0}$ satisfying
\begin{eqnarray}
 \mathbb{P} \left[ \int_{0}^{t} \, \left |r_s \right |  \, {\rm d}s < \infty\right] = 1 ,
\end{eqnarray}
for $t>0$. In that case it follows  that the pricing kernel must be of the form 
\begin{eqnarray} 
\pi_t = \frac{\rho_t} { B_t} \,,
\label{pricing kernel formula}
\end{eqnarray}
where $\{\rho_t\}_{t \geq 0}$ is a strictly positive martingale, and that the price of a generic asset paying no dividend takes the form 
\begin{eqnarray} 
S_t = \frac{B_t \psi_t} { \rho_t} \,,
\label{generic asset}
\end{eqnarray}
where $\{\psi_t\}_{t \geq 0}$ is a martingale. If the asset is of the limited liability type, then $\{\psi_t\}$ is strictly positive. An example of a risky asset that pays no dividend is a foreign money market account, with its value quoted in units of the base currency. 

We are now in a position to introduce a L\'evy-Ito market model driven by an $n$-dimensional Brownian motion and an independent $n$-dimensional Poisson random measure. The market filtration is taken henceforth to be generated by these processes in a standard way. The market consists of a money market account, a pricing kernel, and one or more risky assets, all modelled by  L\'evy-Ito processes.   The short rate is an exogenously specified predictable process and the pricing kernel is assumed to satisfy a dynamical equation of the form
\begin{eqnarray}
{\rm d} \pi_t  = - \pi_{t^{-}} \left[ r_t \, {\rm d}t + \kappa_t  \cdot {\rm d}W_t +  \int_{|x|>0} \Lambda_t(x) \, \tilde{N}({\rm d}x, {\rm d}t)  \right ] . 
\label{pricing kernel dynamics}
\end{eqnarray} 
Here  the predictable vector-valued process $\kappa :  \mathbb R^{+} \times \Omega \rightarrow \mathbb R^n$ can be interpreted as the Brownian market price of risk and is taken to be such that 
\begin{eqnarray} 
\mathbb{P} \left[ \int_{0}^{t}  \kappa_s^{\,2} \, {\rm d}s < \infty \right] = 1 \, .
\label{risk aversion Brownian}
\end{eqnarray}
The predictable process $\{\Lambda_t(x)\}$ can be interpreted as the market price of jump risk for jumps of type $x$ in the $n$-dimensional state space of the Poisson random measure. We assume, with an obvious  slight generalization of the discussion in the previous section to higher dimensions, that 
$\{\Lambda_t(x)\} \in \mathscr{P}_2\{ \mathbb B^n, \mathbb R^+\}$, where $\mathbb B^n$ denotes the interior of the unit ball in $\mathbb R^n$. Thus we have
\begin{eqnarray} 
\mathbb{P} \left[  \int_{0}^{t} \int_{|x| < 1} \Lambda^{\,2}_s(x) \, \nu({\rm d}x) \, {\rm d}s < \infty \right ] = 1\, ,
\label{P_2 condition for Lambda}
\end{eqnarray}
where 
\begin{eqnarray}
|{x}|^2  = \sum_{\alpha=1}^n (x^{\alpha})^2\,.
\end{eqnarray}
To ensure that we can apply Ito's formula, we assume that
\begin{eqnarray} 
\mathbb{P} \left[ \sup_{0 \leq s \leq t} \, \sup_{0 \leq |x |< 1}  \left | \Lambda_s(x)  \right | < \infty \right] = 1
\label{Lambda sup sup 1}
\end{eqnarray}
for $t \geq 0$,  and to ensure that the pricing kernel never drops abruptly to a negative value or to a value that is arbitrarily close to zero we assume that
\begin{eqnarray} 
\mathbb{P} \left[ \sup_{0 \leq s \leq t} \, \sup_{0 \leq |x| < \infty }  \Lambda_s(x) < 1 \right] = 1\,
\label{Lambda sup sup 2}
\end{eqnarray}
for all $t \geq 0$. As a first step towards ensuring that $\{\rho_t\} $ will be martingale we require that
\begin{eqnarray} 
\mathbb{P} \left[ \int_{0}^{t} \int_{|x|\geq 1}  \left | \Lambda_s (x) \right | \, \nu({\rm d}x) \, {\rm d}s < \infty \right] = 1
\label{risk aversion P1 condition 2}
\end{eqnarray}
for all $t \geq 0$. The solution to the stochastic equation for the pricing kernel satisfying these conditions can then be worked out using the methods of Examples 1 and 2, and we obtain
\begin{align} 
\pi_t  =  & \exp \left[  -\int_0^t r_s \, {\rm d}s - \int_{0}^{t}  \kappa_s  \cdot {\rm d}W_s - \tfrac{1}{2}   \int_{0}^{t}  \kappa_s^{\,2} \, {\rm d}s    \right]  \nonumber
\\ 
&
 \times \, \exp \left[- \int_{0}^{t} \int_{|x|>0} \lambda_s (x) \, \tilde{N}({\rm d}x, {\rm d}s) - \int_{0}^{t} \int_{x} \left(\re^{-\lambda_s(x)} -1 + \lambda_s(x) \right) \, \nu({\rm d}x) \, {\rm d}s \right ], 
 \label{Levy Ito pricing kernel RA}
\end{align}
where the map  $\lambda : \mathbb R^n \times \mathbb R^{+} \times \Omega \rightarrow \mathbb R$ appearing here is defined by
\begin{eqnarray}
\lambda_t(x) = - \log \big(1 - \Lambda_t(x) \big)\, .
\label{def of lambda}
\end{eqnarray}
Note that $\lambda_t(x) > 0$ if and only if $\Lambda_t(x) > 0$. We observe that \eqref{Lambda sup sup 1} and \eqref{risk aversion P1 condition 2} ensure that the integrals appearing in \eqref{Levy Ito pricing kernel RA} are well-defined and almost surely finite.
In particular, we find that  $\{\lambda_t(x)\} \in  \mathscr{P}_2\{\mathbb B^n, \mathbb R^+\}$. It also holds that
\begin{eqnarray} 
\mathbb{P} \left[ \int_{0}^{t} \int_{|x|\geq 1}  \re^{-\lambda_s (x)}\, \nu({\rm d}x) \, {\rm d}s < \infty \right] = 1 \, 
\label{lambda condition}
\end{eqnarray}
and
\begin{eqnarray} 
\mathbb{P} \left[ \int_{0}^{t} \int_{|x|\geq 1}  \left | \lambda_s (x) \right | \, \nu({\rm d}x) \, {\rm d}s < \infty \right] = 1
\label{small lambda risk aversion P1 condition 2}
\end{eqnarray}
for all $t \geq 0$.  One can then check that \eqref{Lambda sup sup 2} and \eqref{def of lambda} imply that
\begin{eqnarray} 
\mathbb{P} \left[ \sup_{0 \leq s \leq t} \, \sup_{0 \leq |x| < \infty }  \lambda_s(x) < \infty \right] = 1\,
\label{Lambda sup sup 3}
\end{eqnarray}
for all $t \geq 0$. As a consequence of \eqref{pricing kernel formula}, the dynamical equation satisfied by $\{\rho_t\}$ is 
\begin{eqnarray}
{\rm d} \rho_t  = - \rho_{t^{-}} \left[ \kappa_t  \cdot {\rm d}W_t + \int_{|x|>0} \Lambda_t(x) \, \tilde{N}({\rm d}x, {\rm d}t ) \right], 
\end{eqnarray}
from which it follows that
\begin{align} 
\rho_t  =  & \exp \left[  - \int_{0}^{t}  \kappa_s  \cdot {\rm d}W_s - \tfrac{1}{2}   \int_{0}^{t}  \kappa_s^{\,2} \, {\rm d}s    \right]  \nonumber
\\ 
&
 \times \, \exp \left[- \int_{0}^{t} \int_{|x|>0} \lambda_s (x) \, \tilde{N}({\rm d}x, {\rm d}s) - \int_{0}^{t} \int_{x} \left(\re^{-\lambda_s(x)} -1 + \lambda_s(x) \right) \, \nu({\rm d}x) \, {\rm d}s \right ], 
 \label{rho in exponential form}
\end{align}
and we observe that $\{\rho_t\}$ is a local martingale. Then to impose the martingale property we require
$\mathbb E [ \rho_t ] = 1$ for all $t>0$.  The martingale property imposes further conditions on $\{\kappa_t\}$ and 
$\{\lambda_s (x)\}$ in the initial specification of the dynamics of the pricing kernel at \eqref{pricing kernel dynamics}.

We consider now a typical non-dividend paying limited-liability risky asset with price $\{S_t \}_{t \geq 0}$ in a L\'evy-Ito market with the pricing kernel $\{\pi_t \}$, so 
$\{\pi_t S_t \}_{t \geq 0}$ is a strictly positive martingale.  Writing $\psi_t = \pi_t S_t $ let us assume that the dynamics of $\{\psi_t\}_{t\geq0}$ take the form
\begin{eqnarray}
{\rm d} \psi_t  = \psi_{t^{-}} \left[ \beta_t  \cdot {\rm d}W_t + \int_{|x|>0} \Psi_t(x) \, \tilde{N}({\rm d}x, {\rm d}t ) \right], 
\label{psi dynamics}
\end{eqnarray}
for some predictable  vector-valued process $\{\beta_t \}$ such that 
\begin{eqnarray} 
\mathbb{P} \left[ \int_{0}^{t}  \beta_s^{\,2} \, {\rm d}s < \infty \right] = 1 \, ,
\end{eqnarray}
and for some predictable map-valued process $\{\Psi_t(x)\} \in \mathscr{P}_2\{ \mathbb B^n, \mathbb R^+\}$. We assume that
\begin{eqnarray} 
\mathbb{P} \left[ \sup_{0 \leq s \leq t} \, \sup_{0 \leq |x |< 1}  \left | \Psi_s(x)  \right | < \infty \right] = 1
\label{Psi sup sup 1}
\end{eqnarray}
for $t \geq 0$, and to ensure that $\{\psi_t\}$ never drops abruptly to a negative value or to a value that is arbitrarily close to zero we assume that 
\begin{eqnarray} 
\mathbb{P} \left[ \inf_{0 \leq s \leq t} \, \inf_{0 \leq |x| < \infty }  \Psi_s(x) > - 1 \right] = 1\,
\label{Psi sup sup 2}
\end{eqnarray}
for $t \geq 0$.  To ensure that the dynamics of $\{\psi_t\}$ are well-defined for large jumps, we require 
\begin{eqnarray} 
\mathbb{P} \left[ \int_{0}^{t} \int_{|x|\geq 1}  \left | \Psi_s (x) \right | \, \nu({\rm d}x) \, {\rm d}s < \infty \right] = 1
\label{Psi risk aversion P1 condition 2}
\end{eqnarray}
for $t \geq 0$. With these assumptions it follows that $\{\psi_t\}$ takes the form
\begin{align}
\psi_t = \,& S_0 \, 
 \exp \left[   \int_{0}^{t}  \beta_s  \cdot {\rm d}W_s - \tfrac{1}{2}   \int_{0}^{t}  \beta_s^{\,2} \, {\rm d}s    \right]  \nonumber \\
& \times \exp \left[ \int_{0}^{t} \int_{|x|>0} \beta_s (x) \, \tilde{N}({\rm d}x, {\rm d}s) - \int_{0}^{t} \int_{x} \left(\re^{\beta_s(x)} -1 - \beta_s(x) \right) \, \nu({\rm d}x) \, {\rm d}s  \right].
\end{align}
Here we have used the fact the $\pi_0 = 1$ and we have set
\begin{eqnarray} 
\beta_t (x) = \log \big( 1 + \Psi_t(x) \big)\,.
\end{eqnarray}
We note, in particular, that $\{\beta_t(x)\} \in \mathscr{P}_2\{ \mathbb B^n, \mathbb R^+\}$ and that for all $t \geq 0$ it holds that
\begin{eqnarray} 
\mathbb{P} \left[ \int_{0}^{t} \int_{|x|\geq 1}  \re^{\beta_s (x)}\, \nu({\rm d}x) \, {\rm d}s < \infty \right] = 1 
\label{exp condition on beta}
\end{eqnarray}
and 
\begin{eqnarray} 
\mathbb{P} \left[ \int_{0}^{t} \int_{|x|\geq 1}  \left | \beta_s (x) \right | \, \nu({\rm d}x) \, {\rm d}s < \infty \right] = 1 
\end{eqnarray}
and that for all $t \geq 0$ it holds that
\begin{eqnarray} 
\mathbb{P} \left[ \inf_{0 < s < t} \, \inf_{0 \leq |x| < \infty }  \beta_s(x) > - \infty \right] = 1\,.
\end{eqnarray}
At this stage we are able to work out an explicit expression for the asset price by taking the quotient  \eqref{generic asset}. The result is as follows: 
\begin{align}
S_t = \, & S_0\, B_t \, 
\exp \left[   \int_{0}^{t} ( \beta_s + \kappa_s)  \cdot {\rm d}W_s - \tfrac{1}{2}   \int_{0}^{t}  \beta_s^{\,2} 
+ \tfrac{1}{2}   \int_{0}^{t}  \kappa_s^{\,2} \, {\rm d}s    \right]  
\nonumber \\
&
\times \exp \left[ \int_{0}^{t} \int_{|x|>0} \left( \beta_s(x) + \lambda_s(x)\right)\, \tilde{N}({\rm d}x, {\rm d}s) \right]\nonumber\\
& \, \, \quad \times \exp \left[ - \int_{0}^{t} \int_{x} \left( \re^{\beta_s(x)} -1 - \beta_s(x) \right) \, \nu({\rm d}x) \, {\rm d}s  \right] \nonumber \\
& \, \,   \quad \quad  \times \exp \left[ \int_{0}^{t} \int_{x} \left( \re^{-\lambda_s(x)} -1 + \lambda_s(x) \right) \, \nu({\rm d}x) \, {\rm d}s \right].
\end{align}
Defining $\sigma_t = \beta_t + \kappa_t$ and $\sigma_t(x) = \beta_t(x) + \lambda_t(x)$, we conclude that $\{S_t\}$ takes the form 
\begin{align}
S_t = \, & S_0\, 
\exp \left[  \int_{0}^{t} (r_s + \lambda_s \, \sigma_s )\, {\rm d}s +  \int_{0}^{t}  \sigma_s  \cdot {\rm d}W_s - \tfrac{1}{2}   \int_{0}^{t}  \sigma_s^{\,2} \, {\rm d}s  \right]  
\nonumber \\
\times & \exp \left[ \int_{0}^{t} \int_{|x|>0} \sigma_s\, \tilde{N}({\rm d}x, {\rm d}s) - \int_{0}^{t} \int_{x} \left[ \re^{-\lambda_s(x)}(\re^{\sigma_s(x)} -1) - \sigma_s(x) \right] \, \nu({\rm d}x) \, {\rm d}s \right] .
\label{preliminary form of asset price}
\end{align}
This is the general form an asset price takes in a L\'evy-Ito model. The input processes $\{r_t\}$, 
$\{\lambda_t\}$ and $\{\sigma_t\}$ are unconstrained apart from the conditions required for their definition. 
\vspace{2mm}
\begin{Lemma} 
Writing $\Lambda_t (x) =1 -  \re^{- \lambda_t(x)}$ and $\Sigma_t (x) = \re^{\sigma_t(x)} -1$, we have
\begin{eqnarray} 
\mathbb{P} \left[\int_{0}^{t} \int_{x} \Lambda_t (x) \, \Sigma_t (x) \, \nu({\rm d}x) < \infty \right] = 1.
\label{statement of lemma}
\end{eqnarray} 
\end{Lemma} 
\vspace{2mm}
\noindent \textit{Proof}. We need to show that
\begin{align} 
\mathbb{P} \left[\int_{0}^{t} \int_{|x|<1} \Lambda_t (x) \, \Sigma_t (x) \, \nu({\rm d}x) 
+ \int_{0}^{t} \int_{|x|\geq1} (1 -  \re^{- \lambda_t(x)} ) (\re^{\beta_t(x) + \lambda_t(x)} -1 )\, \nu({\rm d}x)< \infty \right] = 1.
\label{second form of lemma}
\end{align} 
Since $\{\Lambda_t (x)\}$ and $\{\Sigma_t(x) \}$ are in $\mathscr{P}_2\{ \mathbb B^n, \mathbb R^+\}$,  the Cauchy-Schwartz inequality implies that the first term in the square brackets in \eqref{second form of lemma} is finite almost surely. Then as a consequence of  \eqref{Lambda sup sup 3} and \eqref{exp condition on beta} it follows that that the second term is also finite almost surely. \hfill $\square$ 

\vspace{3mm}
\noindent After some simple rearrangement of \eqref{preliminary form of asset price} making use of \eqref{statement of lemma}, we  obtain the following: 
\begin{Proposition} 
The price of a risky asset that pays no dividend in a general L\'evy-Ito model takes the form
\begin{align}
& S_t   =  S_0 \exp \left[ \int_{0}^{t} (r_s + R_s)\, {\rm d}s + \int_{0}^{t}  \sigma_s  \cdot {\rm d}W_s - \tfrac{1}{2}   \int_{0}^{t}  \sigma_s^{\,2} \, {\rm d}s    \right]  \nonumber
\\ 
& \times \exp \left[  \int_{0}^{t} \int_{|x|>0} \log (1 + \Sigma_s (x))  \tilde{N}({\rm d}x, {\rm d}s) - \int_{0}^{t} \int_{x} \left( \Sigma_s(x) - \log (1 + \Sigma_s (x)) \right) \, \nu({\rm d}x) \, {\rm d}s \right] ,
\nonumber
\end{align}
where $\{r_t\}$ is the interest rate, $\{R_t\}$ is the excess rate of return above the interest rate,  and $\{\sigma_t\}$ is the vector Brownian  volatility. The excess rate of return is given by
\begin{align}
R_t = \kappa_t \cdot \sigma_t +  \int_{x} \Lambda_t (x) \, \Sigma_t (x) \, \nu({\rm d}x)\, , 
\label{excess rate of return above the interest rate}
\end{align}
where $\{\kappa_t\}$ is the vector Brownian market price of risk, $\{\Lambda_t (x)\}$ is the market price of jump risk, and 
$\{\Sigma_t (x)\}$ is the dynamical jump volatility.
\end{Proposition} 
\vspace{2mm}

\noindent It may be helpful if we make a few interpretive remarks. First, we observe that the risky asset satisfies the following dynamical equation: 
\begin{eqnarray}
{\rm d}S_t = & S_{t^{-}} \left[\left( r_t + R_t \right)\,  {\rm d}t 
+ \sigma_t \cdot {\rm d}W_t +  \int_{|x|>0} \Sigma_t (x) \, \tilde{N}({\rm d}x, {\rm d}t)  \right].
\end{eqnarray}
The dynamical volatility $\Sigma_t(x)$ represents the riskiness of the asset associated with the point $x$ in the state space of the Poisson random measure at time $t$. Thus, $\Sigma_t(x)$ determines the multiplicative factor by which the price of the asset jumps if the jump in the underlying $n$-dimensional L\'evy process is the vector $x$.
Then $\Lambda_t (x)$ is the market price of risk associated with $x$ at time $t$. The product $\Lambda_t (x)\, \Sigma_t(x) $ is the excess rate of return per unit of jump intensity at $x$, and the L\'evy measure $\nu({\rm d}x)$ determines the jump  intensity. A sufficient condition for the excess rate of return to be strictly positive is that the components of the vector processes $\sigma_t $ and 
$\kappa_t $ are positive for all $t \geq 0$ and that $\Sigma_t(x) > 0$ and
$\Lambda_t (x) > 0$ for all $t \geq 0$ and all $x \in \mathbb R^n$. Then the excess rate of return is an increasing function of both the level of risk and level of risk aversion. Proposition 1 extends analogous results known to hold for models driven by L\'evy processes \cite{Brody Hughston Mackie 2012, Mackie}.

\section{L\'evy-Ito models for Interest Rates}
\label{sec:IRM}
\noindent

\noindent Interest rate models driven by L\'evy processes and other more general processes with jumps have been considered by numerous  authors in the past; see  for example  \cite{Eberlein 1999, Eberlein 2005, Eberlein Ozkan 2005, Bjork 1, Bjork 2, Filipovic,  Biagini, HKT, GS, Brody Hughston 2018, KP} and references cited therein. In what follows we look in detail at interest rate models  of the L\'evy-Ito type. These models are of interest on account of their simplicity, tractability and their suitability for implementation. 

We begin with a few remarks about interest rate modelling in general. There are several different ways of putting together interest rate models, depending on the purpose of the model and on which ingredients of the model one regards as primitives. This accounts for the various  ``approaches" to interest rate theory that have been put forward over the last few decades. But even in the case of a Brownian filtration the relationship of the various modelling frameworks is not easy to summarize in a few words \cite{Hughston, Baxter, Hunt Kennedy, Rogers 1995, Jin Glasserman, Hughston 2003}. 

We take the view that an interest-rate model consists of the following: (i) a pricing kernel $\{\pi_t\}_{t \geq 0}$, (ii) a money market account $\{B_{t}\}_{t \geq 0}$, and (iii) a system of discount bonds $\{P_{tT}\}_{t \geq 0, \,T \geq 0}$ satisfying the relations governing risky assets discussed in the Section III.  A unit discount bond with maturity $T$ pays a dividend of one unit of currency at $T$. Its value drops to zero at  $T$ and stays at that level for all $t > T$. Thus,
\begin{eqnarray}
\lim_{\,t\,\uparrow \, T}P_{t T} = 1
\end{eqnarray}
 and $P_{tT} = 0$ for $ t  \geq T$.
Occasionally, it is useful to refer to the associated discount function 
$\{\bar P_{tT}\}_{0 \leq t \leq T < \infty}$, defined by 
$\bar P_{tT} = P_{tT}$ for $0 \leq t < T < \infty$ and  $\bar P_{TT} = 1$ for $T \geq 0$. The discount function is not defined for $t > T$. One regards $P_{tT}$ as being a price, whereas for each outcome of chance the discount factor  
$\bar P_{tT}$ is a pure number. 

There are five processes that play key roles in the formulation of an interest rate model of the 
L\'evy-Ito type: the short rate $\{r_t\}$, the market price of Brownian risk $\{\kappa_t\}$, the market price of jump risk $\{\Lambda_t(x)\}$, the Brownian volatility structure $\{\omega_{tT}\}$, 
and the jump volatility structure $\{\Omega_{tT}(x)\}$. 
In so-called short-rate models, the short rate and the market price of risk processes are the ``primitives". Once these are specified, the remaining elements of the model can be worked out. In so-called volatility models, which have been popular with practitioners, the discount bond volatility structures and the market price of risk processes are the primitives, and from these we can work out the remaining elements, such as the discount bond prices, the short rate, Libor rates, swap rates, and so on. 

Historically, in a Brownian context, short-rate models were the first to be developed, in the 1970s and 1980s; volatility models came later, in the late 1980s and on into the 1990s, in conjunction with the rise of interest rate derivatives markets. A variant on the volatility model approach, dating from the late 1980s, was to use the instantaneous forward rate volatilities as the primitives, along with the market price of risk \cite{HJM}. The resulting so-called HJM models were highly influential in their day and had a transformative effect on the subject, even though it can be difficult to argue on a practical basis why one would wish to regard the unobservable instantaneous forward rate volatilities as primitives. The so-called Libor market models, dating from the early and mid 1990s, also fall into the volatility model category \cite{BGM}, and these too have been popular, particularly for applications in industry.

A variation on the idea of the short-rate model also dating from the early and mid 1990s was that of combining the short rate and the market price of risk together to form a pricing kernel (or state-price density), and using that as the primitive \cite{Constantinides, Flesaker Hughston 1996, Rogers 1997, Jin Glasserman}. 
From a broad perspective, short-rate models, volatility models, and pricing kernel models are more or less equivalent, modulo technicalities. Where they differ is in the ease with which specific models can be developed, and in the facility with which parametric and functional degrees of freedom can be incorporated in order to calibrate the models to market data.

When it comes to the formulation of L\'evy-Ito models for interest rates, it will be convenient to begin with the volatility approach. This is because the ideas that we have developed in the previous section concerning risky assets can be carried over directly. 
We regard the discount bond volatility structures as being given, along with the associated market prices of risk. Following the scheme outlined in the previous section, including specification of the pricing kernel according to equation \eqref{Levy Ito pricing kernel RA}, we treat each discount bond as a risky asset, and for the dynamics of a $T$-maturity  bond we write
\begin{align}
{\rm d}P_{tT} = P^{-}_{tT} \left[\left( r_t + \kappa_t \cdot \omega_{tT} + \int_{x} \Lambda_t (x)  \, \Omega_{tT} (x)\,  \nu({\rm d}x) \right)\,  {\rm d}t + \omega_{t T} \cdot {\rm d}W_t +  \int_{|x|>0} \Omega_{tT} (x) \, \tilde{N}({\rm d}x, {\rm d}t)  \right]. 
\nonumber \\
\end{align}
Here for convenience we write 
\begin{eqnarray}
P^{-}_{tT} = \lim_{\,s\, \uparrow \,t}P_{sT}.
\end{eqnarray}
The discount bond Brownian vector volatility structure and Poisson jump volatility structure are denoted  $\{\omega_{tT}\}$ and  $\{\Omega_{tT}(x)\}$, respectively. Then as a consequence of Proposition 1 we deduce that the discount bond system takes the form
\begin{align}
P_{tT}  =  & \,  \mathds{1}(t<T)\,P_{0T}\,\exp \left[\int_0^t (r_s + R_{sT} )\, {\rm d}s 
+ \int_{0}^{t} \omega_{sT}  \cdot {\rm d}W_s + \int_{0}^{t} \int_{|x|>0} \log(1 + \Omega_{sT}(x)) \, \tilde{N}({\rm d}x, {\rm d}s) \right]
 \nonumber\\ & \times  \exp \left[ -\half \int_{0}^{t}  \,  \omega_{sT}^{\,2} \, {\rm d}s -\int_{0}^{t} \int_{x}   \left(\Omega_{sT}(x) -\log (1 + \Omega_{sT}(x))\right) \, \nu({\rm d}x) \, {\rm d}s  \right],
\label{Discount Bond Price Formula}
\end{align} 
where the excess rate of return is given for $t<T$ by
\begin{eqnarray}
R_{tT} =   \kappa_t \cdot \omega_{tT} + \int_{x} \Lambda_{t}(x) \, \Omega_{tT}(x) \,  \nu({\rm d}x) \, .
\label{bond excess rate of return}
\end{eqnarray}
We require that the volatility structures  satisfy 
\begin{eqnarray}
\lim_{\, t\, \uparrow\,T}\omega_{t T} = 0, \quad \lim_{\, t\, \uparrow\,T}\Omega_{t T} = 0 
\end{eqnarray}
for each $T>0$.
It follows then from the maturity condition on the discount bond and the definition of the unit-initialized money market account that we have the following: 
\vspace{3mm}
\begin{Proposition}
\label{Ch3 P2}
In a L\'evy-Ito interest rate model, let the vector market price of Brownian risk $\{\kappa_t\}$, the vector Brownian volatility structure  $\{\omega_{tT}\}$, the market price of jump risk 
$\{\Lambda_t(x)\}$, and the jump volatility structure $\{\Omega_{tT}(x)\}$ be given, along with the initial term structure $\{P_{0t}\}_{t>0}$. Then the money market account takes the form 
\begin{align}
B_t = \,& (P_{0t})^{-1}\,\exp \left[-\int_0^t R_{st} \, {\rm d}s - \int_{0}^{t} \omega_{st}  \cdot {\rm d}W_s -\int_{0}^{t} \int_{|x|>0} \log(1 + \Omega_{st}(x)) \tilde{N}({\rm d}x, {\rm d}s) \right] \nonumber\\
& \times \exp \left[ \tfrac{1}{2} \int_{0}^{t}  \omega_{st}^{\,2} \, {\rm d}s + \int_{0}^{t} \int_{x}  \left(\Omega_{st}(x) -\log(1 + \Omega_{st}(x))\right) \, \nu({\rm d}x) \, {\rm d}s  \right].
\label{Money Market Account}
\end{align}
where $\{R_{st}\}$ is given by \eqref{bond excess rate of return}.
\end{Proposition}
\vspace{3mm}
\noindent Substituting (\ref{Money Market Account}) into (\ref{Discount Bond Price Formula}), we obtain the following general representation of interest rates in a L\'evy-Ito setting: 
\vspace{3mm}
\begin{Proposition}
In a L\'evy-Ito interest rate model, let the vector market price of Brownian risk $\{\kappa_t\}$, the vector Brownian volatility structure  $\{\omega_{tT}\}$, the market price of jump risk 
$\{\Lambda_t(x)\}$, and the jump volatility structure $\{\Omega_{tT}(x)\}$ be given, along with the initial term structure $\{P_{0t}\}_{t>0}$. Then the price of a unit discount bond with maturity $T$ takes the form 
\begin{align}
 P_{tT} & = \mathds{1}(t<T)\,P_{0tT}\,
 \exp \left[\int_0^t  \kappa_{t} \cdot (   \omega_{sT} - \omega_{st}) \, {\rm d}s + \int_0^t
  \int_{x}\Lambda_{s}(x) \, (   \Omega_{sT}(x) - \Omega_{st}(x)) \, \nu({\rm d}x) \, {\rm d}s \,\right] \nonumber \\
  & \hspace{1cm}\times \exp \left[ \int_{0}^{t}   (   \omega_{sT} - \omega_{st}) \cdot {\rm d}W_s - \tfrac{1}{2} \int_{0}^{t} (\omega_{sT}^{\,2} - \omega_{st}^{\,2}) \,{\rm d}s \right]\nonumber \\
 & \hspace{1cm}\times \exp \left[ \int_{0}^{t} \int_{|x|>0}  \log \left( \frac{1 + \Omega_{sT}(x)} {1 + \Omega_{st}(x)} \right) \, \tilde{N}({\rm d}x, {\rm d}s) \right]\nonumber \\
 & \hspace{1cm}\times \exp \left[- \int_{0}^{t} \int_{x}  \left(\Omega_{sT}(x) -\Omega_{st}(x) - \log \left( \frac{1 + \Omega_{sT}(x)} {1 + \Omega_{st}(x)} \right) \right) \, \nu({\rm d}x) \, {\rm d}s  \right], 
\label{Discount Bond Price Formula Final}
\end{align}
where $P_{0tT} = P_{0T}/P_{0t}$ denotes the forward price made at time $0$ for purchase at time $t$ of a unit
$T$-maturity bond.
\end{Proposition}
\vspace{3mm}
Thus we see that once the initial term structure, the market price of risk processes, and the volatility structures have been specified, the money market account and the discount bond prices are determined. To propose a specific interest rate model one needs to choose a parametric form for the market prices of risk and the volatility structures sufficiently general to allow one to calibrate the model to the initial term structure and to an appropriate range of liquid market prices for interest rate options, futures contracts,  and other derivatives. The resulting fully calibrated model can then be used both in simulation studies for risk management and investment analysis, as well as for pricing and trading complex derivatives.  The L\'evy-Ito models have an advantage over  L\'evy models and Brownian models in that the functional freedom available for calibration in the L\'evy-Ito case is much more flexible. 
\section{Vasicek model of the L\'evy-Ito type}
\label{sec:LIVM}
\noindent

\noindent As a non-trivial example of a L\'evy-Ito interest rate model derived via the short-rate method, we construct a Vasicek model of the 
L\'evy-Ito type. In the L\'evy-Ito Vasicek model,  the short rate $\{r_t\}_{t \geq 0}$ is taken to be a mean-reverting 
process of the Ornstein-Uhlenbeck (OU) type, satisfying a stochastic differential equation of the form
\begin{eqnarray}
\rd r_t = k(\theta - r_t) \, \rd t -  \kappa(t)\,{\rm d} W_t -  \int_{|x|>0} \sigma (x, t) \, \tilde{N}({\rm d}x, {\rm d}t) \,  ,
\label{OU dynamics}
\end{eqnarray} 
where $x \in \mathbb{R}^n$. The strictly positive constants $k$ and $\theta$ denote the mean reversion rate and the mean reversion level. We assume that the deterministic function $\kappa :\mathbb{R}^+ \rightarrow \mathbb{R}^+$ satisfies 
\begin{eqnarray}
 \int_0^t  \kappa^2(s) \,{\rm d}s < \infty\, ,
\end{eqnarray} 
for all $t>0$, and that the  left-continuous deterministic function $\sigma :\mathbb{R}^n \times \mathbb{R}^+  \rightarrow \mathbb{R}^+$ satisfies
\begin{eqnarray} 
 \int_0^t\int_{|x|<1} \sigma^2(x,s) \, \nu({\rm d}x) \,{\rm d}s < \infty \, , \quad \int_0^t \int_{|x|\geq 1} \exp \left[\frac{1}{k}{\sigma(x,s)}\right]  \nu({\rm d}x) \,{\rm d}s < \infty \,,
 \label{conditions on Vasicek sigma}
\end{eqnarray}
for all $t>0$. The initial value of the short rate is $r_0$ and the initial value of the money market account is unity. In a more general version of the model we could also let $k$ and $\theta$ be functions of time, but for simplicity we keep these parameters constant. To ease the notation further going forward, we omit the Brownian term. This term can be easily restored.  The risk aversion process is taken to be a  left-continuous deterministic function $\lambda :\mathbb{R}^n \times \mathbb{R}^+ \rightarrow \mathbb{R}^+$ chosen such that 
\begin{eqnarray} 
 \int_0^t \int_{|x|<1} \lambda^2(x, s) \, \nu({\rm d}x) \,{\rm d}s< \infty \, ,\quad  \int_0^t \int_{|x|\geq 1}  \lambda(x,s) \, \nu({\rm d}x) \,{\rm d}s< \infty \,,
  \label{conditions on Vasicek lambda}
\end{eqnarray}
for $t>0$. Then the strictly positive process $\{M_t\}_{t \geq 0}$ defined by
\begin{eqnarray}
M_t = \exp \left[-\int_{0}^{t} \int_{|x|>0} \lambda(x,s) \, \tilde{N}({\rm d}x, {\rm d}s) - \int_{0}^{t} \int_{x} \left(\re^{-\lambda(x,s)} -1 + \lambda(x,s) \right) \, \nu({\rm d}x) \, {\rm d}s \,  \right]
\label{Risk aversion vasicek martingale}
\end{eqnarray} 
is a martingale, and the stochastic differential equation (\ref{OU dynamics}) can be solved to give 
\begin{eqnarray}
r_t = \theta + (r_0 - \theta)\, \re^{-kt} -\int_{0}^{t} \int_{|x|>0} \re ^ {k(s-t)} \sigma (x,s) \, \tilde{N}({\rm d}x, {\rm d}s) \, . 
\label{short rate}
\end{eqnarray}
We observe that the mean of $r_t$ is $\theta + (r_0 -\theta)\, \re^{-kt}$ and that for the variance we have 
\begin{eqnarray}
{\rm Var}\,[r_t] =  \int_{0}^{t}\int_{x} \re ^ {2k(s-t)} \sigma^2(x,s) \, \nu({\rm d} x) \,{\rm d}s\,. 
\end{eqnarray}

To obtain explicit formulae for the money market account \eqref{money market account formula} and the pricing kernel \eqref{pricing kernel formula}, we require an expression for the integrated 
short rate, 
\begin{eqnarray}
I_t = \int_0^t r_s \,\rd s \, .
\label{integrated short rate}
\end{eqnarray}
\noindent This can be obtained by integrating \eqref{OU dynamics} directly and rearranging the result. We get
\begin{eqnarray} 
I_t = \theta t - \frac{1}{k} (r_t - r_0) - \frac{1}{k}\int_{0}^{t} \int_{|x|>0} \sigma (x,s) \, \tilde{N}({\rm d}x, {\rm d}s) \, .  
\end{eqnarray}
It follows that the money market account is given by 
\begin{eqnarray} 
B_t =  \exp \left[ \theta t  -  \frac{1}{k} (r_t - r_0)  - \frac{1}{k} \int_{0}^{t}\int_{|x|>0} \sigma (x,s) \, \tilde{N}({\rm d}x, {\rm d}s) \right] , 
\end{eqnarray}
and that the pricing kernel can be expressed in the form 
\begin{align} 
& \pi_t =  \exp \left[ - \theta t + \frac{1}{k} (r_t - r_0)\right] 
\exp \left[\int_{0}^{t} \int_{|x|>0} \left(\frac{1}{k} \sigma(x,s)-\lambda (x,s) \right)\, \tilde{N}({\rm d}x, {\rm d}s) \right] \nonumber
\\
& \hspace{2.13cm} \times \exp \left[- \int_{0}^{t} \int_{x} \left(\re^{-\lambda(x,s)} -1 + \lambda(x,s) \right) \, \nu({\rm d}x) \, {\rm d}s \right]. 
\label{Vasicek pricing kernel with short rate} 
\end{align}
Armed with these expressions, we proceed to derive an expression for the price of a discount bond using the well-known  discount-bond valuation formula of Constantinides \cite{Constantinides} :
\begin{eqnarray}
P_{tT} = \mathds{1}\{t<T\} \,\frac{1}{\pi_t} \, \mathbb{E}_t[\pi_T] \, .
\label{Constantinides formula}
\end{eqnarray}
The conditional expectation of $\pi_T$ is given by
\begin{align}
\mathbb E_t [ \pi_T] & =  \exp \left[- \theta T  - \frac{1}{k} \left(1 - {\rm e}^{-kT}\right)  (r_0 - \theta) \right] \nonumber \\ 
&  \times \exp \left[- \int_{0}^{T} \int_{x} \left(\re^{-\lambda(x,s)} -1 + \lambda(x,s) \right) \, \nu({\rm d}x) \, {\rm d}s  \right] \nonumber \\   
&  \times \exp \left[\int_{0}^{t} \int_{|x|>0} \left(\frac{1}{k}   \left(1 - \re^{k(s-T)}\right) \sigma(x,s)-\lambda (x,s) \right) \tilde{N}({\rm d}x, {\rm d}s) \right] \nonumber
\\ 
&  \times \mathbb E \left[ \exp \int_{t}^{T} \int_{|x|>0}  \left(\frac{1}{k}   \left(1 - \re^{k(s-T)}\right) \sigma(x,s)-\lambda (x,s) \right) \tilde{N}({\rm d}x, {\rm d}s) \right],
\end{align}
for $t < T$, which if we introduce the short rate takes the simpler form
\begin{align}
\mathbb E_t [ \pi_T] & =  \exp \left[- \theta T - \frac{1}{k} (r_0 - \theta) +
\frac{1}{k} {\rm e}^{-k(T-t)}  (r_t - \theta) \right] \nonumber \\ 
&  \times \exp \left[- \int_{0}^{T} \int_{x} \left(\re^{-\lambda(x,s)} -1 + \lambda(x,s) \right) \, \nu({\rm d}x) \, {\rm d}s  \right] \nonumber \\   
&  \times \exp \left[\int_{0}^{t} \int_{|x|>0} \left(\frac{1}{k}  \sigma(x,s)-\lambda (x,s) \right) \tilde{N}({\rm d}x, {\rm d}s) \right] \nonumber
\\ 
&  \times \mathbb E \left[ \exp \int_{t}^{T} \int_{|x|>0}  \left(\frac{1}{k}   \left(1 - \re^{k(s-T)}\right) \sigma(x,s)-\lambda (x,s) \right) \tilde{N}({\rm d}x, {\rm d}s) \right].
\end{align}
Then dividing by our expression \eqref{Vasicek pricing kernel with short rate} for the pricing kernel, one sees that there is some cancelation and we obtain
\begin{align}
\frac{1}{\pi_t} \mathbb E_t [ \pi_T] & =  \exp \left[- \theta (T-t) - 
\frac{1}{k} \left(1 - {\rm e}^{-k(T-t)}\right)  (r_t - \theta) \right] \nonumber \\ 
&  \times \exp \left[- \int_{t}^{T} \int_{x} \left(\re^{-\lambda(x,s)} -1 + \lambda(x,s) \right) \, \nu({\rm d}x) \, {\rm d}s  \right] \nonumber \\   
&  \times \mathbb E \left[ \exp \int_{t}^{T} \int_{|x|>0}  \left(\frac{1}{k}   \left(1 - \re^{k(s-T)}\right) \sigma(x,s)-\lambda (x,s) \right) \tilde{N}({\rm d}x, {\rm d}s) \right].
\label{conditional expectation of PK}
\end{align}
It remains to work out the expectation in the third term. Now, for any deterministic left-continuous process $\{f(x,t)\}_{t \geq 0, \, x \in \mathbb{R}^n }$ satisfying
\begin{eqnarray} 
 \int_{0}^{t} \int_{|x|<1} f^2 (x,s) \, \nu({\rm d}x)\, {\rm d}s < \infty
\end{eqnarray}
and
\begin{eqnarray} 
 \int_{0}^{t} \int_{|x|\geq 1} \re^{f(x,s)}  \, \nu({\rm d}x) \, {\rm d}s < \infty\, , \quad
 \int_{0}^{t} \int_{|x|\geq 1} \left | f(x,s) \right |  \, \nu({\rm d}x) \, {\rm d}s < \infty \, ,
\end{eqnarray}
for $t\geq 0$, we can make use of the so-called exponential formula
\begin{eqnarray}
  \mathbb{E}_t   \left[\exp  \int_t^T \int_{|x|>0} f (x,s)\, \tilde{N}({\rm d}x, {\rm d}s) \right] = \exp \left[\int_{t}^{T} \int_{x} \left(\re^{f(x,s)} -1 - f(x,s) \right) \, \nu({\rm d}x) \, {\rm d}s\right] .  
  \label{exponential property}
\end{eqnarray}
Therefore, if for each fixed $T>0$ we define
\begin{eqnarray} 
f_T(x,s) =
\frac{1}{k}   \left(1 - \re^{k(s-T)}\right) \sigma(x)-\lambda (x) \, 
\label{alpha_T}
\end{eqnarray}
for $0 \leq s \leq T$, then by  \eqref{conditions on Vasicek sigma} and   \eqref{conditions on Vasicek lambda} we obtain
\begin{align}
  \mathbb{E}_t \left[  \exp \left[ \int_t^T \int_{|x|>0} f_T(x,s)\, \tilde{N}({\rm d}x, {\rm d}s) \right] \right] = \exp \left[\int_{t}^{T} \int_{x} \left(\re^{f_T(x,s)} -1 - f_T(x,s) \right) \, \nu({\rm d}x) \, {\rm d}s  \right]. \nonumber
  \\
  \label{exponential property with alpha_T}
\end{align}
Finally, using  \eqref{conditional expectation of PK}, \eqref{alpha_T}, and  \eqref{exponential property with alpha_T}, we arrive at the following\,:
\vspace{2mm}
\begin{Proposition} \, In a L\'evy-Ito interest rate model for which the short rate of interest \{$r_t\}$ satisfies an Ornstein-Uhlenbeck equation with mean reversion rate $k$, with mean reversion level $\theta$, with deterministic jump risk aversion $\{\lambda(x,t)\}$, and with deterministic jump volatility $\{\sigma(x,t)\}$, the discount bond system is given for $0 \leq t < T$ by
\begin{align}
 & \log P_{tT}   = 
   \, -(T - t)\, \theta 
  \, + \, \frac{1}{k} \left(1 -\re^{k(t-T)}\right) (\theta - r_t) \nonumber\\
  & \quad + \int_t^T \int_{x}  \left [  \left(   \exp  \left[ \frac{1}{k} \left(1 -\re^{k(s-T)}\right)   \sigma(x,s) \right]  -1 \right ) \re^{-\lambda(x,s)} 
  \, - \frac{1}{k} \left(1 -\re^{k(s-T)}\right)   \sigma(x,s) \right ]  \, \nu({\rm d}x ) \, {\rm d}s \, .
 \label{DB_BM}
\end{align}
\end{Proposition}
\vspace{2mm}
\noindent Thus, by use of a pricing kernel technique we have obtained an expression for the price of a unit discount bond of maturity $T$ in the L\'evy-Ito Vasicek model, generalizing results of  \cite{Norberg 2004, Brody Hughston Meier 2018, Vasicek, Cairns}. The extra freedom provided by the functions $\{\lambda(x,t)\}$ and $\{\sigma(x,t)\}$ gives the model flexibility when it comes to fitting it to market data. One of the interesting features of the L\'evy-Ito Vasicek model is that by allowing risk aversion to vary as a function of jump size and time one can let agents be, for example, much more risk-averse to negative jumps than to positive jumps, while allowing for the possibility that the degree of risk aversion for various jump levels may change as time passes. Such behavioral characteristics can be accommodated into L\'evy-Ito models.  

The so-called L\'evy-Vasicek models \cite{Norberg 2004, Brody Hughston Meier 2018, Eberlein Kallsen}, obtained by setting $n = 1$ with $\lambda(x,t) = \lambda x$ and $\sigma(x,t) = \sigma x$ for $\lambda, \sigma \in \mathbb R$, constitute a more specialized class, and do not allow for the possibility of showing different levels of risk aversion or volatility for different jump sizes and time frames. In a L\'evy-Vasicek model, once we reinstate the Brownian term, the dynamical equation satisfied by the short rate takes the form  
\begin{eqnarray}
\rd r_t = k(\theta - r_t) \, \rd t - \kappa \,{\rm d}W_t - \sigma \, {\rm d}\xi_t \,, \quad
\xi_t = \int_{0}^{t} \int_{|x|>0} x \, \tilde{N}({\rm d}x, {\rm d}s) \, .
\end{eqnarray} 
It should be evident that the L\'evy-Ito Vasicek model can be generalized by restoring the Brownian term and incorporating a deterministic time dependence in the mean reversion rate and the mean reversion level in the spirit of \cite{Hull White, Jamshidian}. \\

\section{L\'evy-Ito Chaos Models for Interest Rates}
\label{sec:CM}
\noindent
The rather broad class of  L\'evy-Ito interest rate models that we investigate in this section arises as an example of the use of  the pricing kernel method and has the defining property that the pricing kernel can be expressed as the conditional variance of an $\mathscr{F_{\infty}}$-measurable square-integrable random variable. Now, it is well known (\cite{Meyer}, page 164) that the conditional variance of a square-integrable random variable on a filtered probability space is a potential of type $D$. This property makes such a process a viable candidate for consideration as a pricing kernel \cite{Flesaker Hughston 1996, Rogers 1997}. We shall be concerned in what follows with the more subtle matter of determining the class of models for which the pricing kernel admits a conditional variance representation. It is known that a large class of Brownian interest rate models can be constructed with this property \cite{Rafailidis, Hughston Rafailidis 2005, Brody Hughston 2004}. Here we address the construction of a more general class of conditional variance models, incorporating jumps.

The setup is is follows. We consider the L\'evy-Ito model for the pricing kernel $\{\pi_t\}_{t \geq 0}$ formulated as in Section III.  For simplicity we suppress the Brownian dependence of this process. We assume that the interest rate $\{r_t\}_{t \geq 0}$ is strictly positive and that the model supports the existence of a floating rate note paying the short rate of interest on a unit principal on a continuous basis in perpetuity. The value of such a note is unity. Thus, we have $S_t = 1$ for all $t \geq 0$ and $\Delta ({\rm d}t) = r_t  {\rm d}t$, and  by the valuation formula \eqref{value from dividends} we have
\begin{eqnarray}
1 = \frac{1}{\pi_t} \mathbb{E}_t\left[ \int_t^{\infty} r_s \pi_s  \, {\rm d}s\right].
\label{FRN valuation formula}
\end{eqnarray}
The intuition behind the pricing formula is that if interest is paid on a unit principal on a continuous basis, then the account will accumulate in value on an exponential basis---this leads to the standard expression for a continuous money market account. 
But if the interest is paid out continuously as a dividend, then the account itself must remain constant in value, and we are led to \eqref{FRN valuation formula}. 
It follows from the foregoing considerations that the pricing kernel can be expressed as a conditional expectation of the form
\begin{eqnarray}
\pi_t = \mathbb{E}_t\left[ \int_t^{\infty} r_s \pi_s  \, {\rm d}s\right].
\label{pricing kernel consistency}
\end{eqnarray}
Such a relation holds in any pricing kernel model with positive interest admitting a floating rate note \cite{Hughston Rafailidis 2005, Brody Hughston 2004, Rafailidis}. In particular, we can take as our basic assumption the condition that 
\begin{eqnarray}
 \mathbb{E}\left[ \int_0^{\infty} r_s \pi_s  \, {\rm d}s\right] < \infty \, ,
 \label{integrability}
\end{eqnarray}
where the integrand is strictly positive. 
Now suppose that a process $\{\gamma_t(x)\}$ is predictable and satisfies 
\begin{eqnarray}
 \mathbb{P}\left[ \int_0^{t} \int_x \gamma_s(x)^2 \,\nu({\rm d}x)\, {\rm d}s < \infty \right] = 1\, . 
 \label{P2 extended}
\end{eqnarray}
Then the process $\{X_t\}$ defined by
\begin{eqnarray}
X_t = \int_{0}^{t} \int_{|x|>0} \gamma_s(x) \, \tilde{N}({\rm d}x, {\rm d}s) 
\end{eqnarray}
is a local martingale.  If additionally
\begin{eqnarray}
 \mathbb{E}\left[ \int_0^{t} \int_{x} \gamma_s(x)^2 \,\nu({\rm d}x)\, {\rm d}s \right] < \infty \,  
 \label{square integrability}
\end{eqnarray}
for all $t\geq0$, then $\{X_t\}$ is a square-integrable martingale with mean zero, and we have
\begin{eqnarray}
 \mathbb{E}\left[ X_t^{\,2} \right] =  \mathbb{E}\left[ \int_0^{t} \int_{x} \gamma_s(x)^2 \,\nu({\rm d}x)\, {\rm d}s \right] .
\end{eqnarray}
Then if
\begin{eqnarray}
 \mathbb{E}\left[ \int_0^{\infty} \int_{x} \gamma_s(x)^2 \,\nu({\rm d}x)\, {\rm d}s \right] < \infty \, . 
 \label{monotone square integrability}
\end{eqnarray}
we find that the limit $X_{\infty} = \lim_{t \to \infty} X_t $ exists almost surely, that  the random variable $X_{\infty}$ is square-integrable, and that the martingale $\{X_t\}$ is closed by $X_{\infty}$. 
Thus we conclude that if a predictable process  $\{\gamma_t(x)\}$ is such that
 \eqref{monotone square integrability} holds, then the integral  
\begin{eqnarray}
X_{\infty} = \int_{0}^{\infty} \int_{|x|>0} \gamma_s(x) \, \tilde{N}({\rm d}x, {\rm d}s) 
\end{eqnarray}
is well-posed and 
\begin{eqnarray}
\mathbb{E}\left[ X_{\infty}^{\,2} \right] < \infty \,.
\end{eqnarray}

Returning to the construction of a pricing kernel, we consider a special case of the preceding, by observing that the predictable process defined by
\begin{eqnarray}
\gamma_t(x) = \frac{1 \wedge |{x}|} {\sqrt{\int_{x} \, (1 \wedge |{x}|^2 )\,\nu({\rm d}x)}} \,\sqrt{r_{s^-} \pi_{s^-}} 
\end{eqnarray}
satisfies \eqref{monotone square integrability} as a consequence of  \eqref{integrability}. 
It is then evident that the random variable 
\begin{eqnarray}
X_{\infty} = \frac{1}{\sqrt{\int_{x} (1 \wedge |{x}|^2 ) \,\nu( {\rm d} x)}} \,  \int_{0}^{\infty} \int_{|x|>0}
{\sqrt{r_{s^-} \pi_{s-}}\,(1 \wedge |{x}| ) \, \tilde{N}({\rm d}x,{\rm d}s)} \, ,
\label{CHI}
\end{eqnarray}
is $\mathscr{F_{\infty}}$-measurable and square-integrable, where
\begin{eqnarray}
\mathscr{F}_{\infty} = \sigma \left( \bigcup_{\,0 \leq t < \infty} \! \mathscr{F}_{t} \right) .
\end{eqnarray}
We proceed to calculate the conditional variance of $X_{\infty}$, which is defined by 
\begin{eqnarray}
\text{Var}_t[X_{\infty}] = \mathbb{E}_t\left[(X_{\infty} - \mathbb{E}_t[X_{\infty}])^2 \right].
\label{CVAR}
\end{eqnarray}
To work out \eqref{CVAR}, we use the conditional Ito isometry for Poisson random measure to obtain
\begin{eqnarray}
\mathbb{E}_t \left[ \left(\int_{t}^{\infty} \int_{|x|>0} \gamma_s(x) \, \tilde{N}( {\rm d}x, {\rm d}s)\right)^2 \right] = \mathbb{E}_t \left[ \int_{t}^{\infty} \int_{x} \gamma^{\,2}_s(x)\, \nu({\rm d}x) \, {\rm d}s \right] ,
\label{LI Isometry}
\end{eqnarray}
which holds under  \eqref{monotone square integrability}.
A calculation making use of  \eqref{CHI}, \eqref{CVAR} and \eqref{LI Isometry} then gives
\begin{eqnarray}
\text{Var}_t \, [X_{\infty}] =  \mathbb{E}_t\left[ \int_t^{\infty} r_s \pi_s  \, {\rm d}s\right] ,
\label{CVAR PK}
\end{eqnarray}
and we see that the conditional variance of the random variable \eqref{CHI} is a pricing kernel of the form \eqref{pricing kernel consistency}.
Thus, we have established the following: 

\vspace{2mm}
\begin{Proposition} 
In any positive interest rate model driven by the Poisson random measure associated with an $n$-dimensional L\'evy process and supporting the existence of a continuous floating rate note, the pricing kernel can be expressed as the conditional variance of a square-integrable $\mathscr{F_{\infty}}$-measurable random variable. 
\end{Proposition}
\vspace{2mm}

We refer to the setup just described as a conditional variance representation of the pricing kernel. This leads us to non-trivial extensions of results obtained in the Brownian case in \cite{Hughston Rafailidis 2005, Brody Hughston 2004, Rafailidis, Grasselli Hurd 2005, Tsujimoto 2010, Grasselli Tsujimoto 2011}, which we now proceed to discuss.

It is well known that in the case of a probability space equipped with the filtration generated by a standard Brownian motion in $n$ dimensions any square-integrable $\mathscr{F_{\infty}}$-measurable random variable  admits a so-called Wiener chaos expansion \cite{Wiener 1938, Ito 1951}. The chaos expansion expresses the random variable in the form of a convergent sum of multiple stochastic integrals, where the $k$-th term involves an integrand given by a function of $k$ time variables defined on a triangular domain, satisfying a square-integrability condition.  

This property extends to the case when the filtration is generated by a Poisson random measure in $n$ dimensions \cite{Ito 1956, Nualart Schoutens 2000, Lokka 2005}, in which case the $k$-th term of the chaos expansion involves an integrand given by a function of $k$ time variables and $k$ space variables, each such space integration being over a copy of the state space of the Poisson random measure. As a consequence, the random variable $X_{\infty}$ associated with the pricing kernel in any interest rate model of the L\'evy-Ito type admits a chaos expansion, providing that the model supports a floating rate note that pays out interest on a continuous basis in perpetuity. If the chaos expansion admits terms only up to order $j$, then we say that we have a general $j$-th order chaos model. If the expansion consists exclusively of the term of order $j$, then we say that we have a pure $j$-th order chaos model. 

As an example of the resulting scheme, we shall present the form taken by the discount bonds in a general second-order chaos model driven by Poisson random measure. In this case we are given a pair of deterministic functions 
\begin{eqnarray}
\{ \phi_s(x)\}_{0\leq s < \infty, \, x \in \mathbb R^n}\, , \quad 
\{\phi_{s\,s_1}(x,x_1)\}_{0\leq s_1 \leq s < \infty, \,  x \in \mathbb R^n, \, x_1 \in \mathbb R^n} 
\label{chaos coefficients constraints}
\end{eqnarray}
satisfying
\begin{eqnarray}
\int_{s=0}^\infty \int_{x}\phi^{\,2}_s(x)\,\nu({\rm d}x) \, {\rm d}s < \infty \,, 
\end{eqnarray}
and
\begin{eqnarray}
 \int_{s=0}^\infty \int_{x}\int_{s_1= 0}^{s}\int_{x_1}\phi^{\,2}_{s\,s_1}(x,x_1)\, \nu({\rm d}x_1) \, {\rm d}s_1\, \nu({\rm d}x) \, {\rm d}s < \infty \,.
\end{eqnarray}
These two functions are used to define an $\mathscr{F}_{\infty}$-measurable random variable given by
\begin{align}
X_\infty = \int_{s=0}^\infty \int_{|x|>0}\phi_s(x) \, \tilde{N}( {\rm d}x, {\rm d}s) + \int_{s=0}^\infty \int_{|x|>0}\int_{s_1 = 0}^{s}\int_{|x_1|>0}\phi_{s\,s_1}(x,x_1) \, \tilde{N}( {\rm d}x_1, {\rm d}s_1) \, \tilde{N}( {\rm d}x, {\rm d}s) , \,\,\,
\label{Wiener Expansion}
\end{align}
where we have $x \in \mathbb R^n$ and $x_1 \in \mathbb R^n$. The first step in the determination of the associated interest rate model is to calculate the conditional variance of $X_\infty$.  The pricing kernel is
\begin{eqnarray}
\pi_t = \mathbb{E}_t\left[(X_{\infty} - \mathbb{E}_t[X_{\infty}])^2 \right],
\end{eqnarray}
A calculation that includes breaking the time integration and making use of the independent increments property gives
\begin{align}
I_{t} =\int_{s=t}^\infty \int_{|x|>0}\phi_s(x) \, \tilde{N}( {\rm d}x, {\rm d}s) + \int_{s=t}^\infty \int_{|x|>0}\int_{s_1 = 0}^{s}\int_{|x_1|>0}\phi_{s\,s_1}(x,x_1) \, \tilde{N}( {\rm d}x_1, {\rm d}s_1) \, \tilde{N}( {\rm d}x, {\rm d}s)\,,
\end{align}
where $I_{t} = X_{\infty} - \mathbb{E}_t[X_{\infty}]$. Then we break the $s_1$ integration at $t$ to obtain
\begin{align}
I_{t} &=\int_{s=t}^\infty \int_{|x|>0}\left(\phi_s(x)  +\int_{s_1 = 0}^{t}\int_{|x_1|>0}\phi_{s\,s_1}(x,x_1) \, \tilde{N}( {\rm d}x_1, {\rm d}s_1) \right)\, \tilde{N}( {\rm d}x, {\rm d}s) \nonumber \\
& + \int_{s=t}^\infty \int_{|x|>0}\int_{s_1 = t}^{s}\int_{|x_1|>0}\phi_{s\,s_1}(x,x_1) \, \tilde{N}( {\rm d}x_1, {\rm d}s_1) \, \tilde{N}( {\rm d}x, {\rm d}s) \, .
\label{Giota}
\end{align}
Forming the square of \eqref{Giota} we get
\begin{eqnarray}
I^2_{t} &=  \zeta_{t} + \eta_{t} + 2 \theta_{t}
\label{Giota^2}
\end{eqnarray}
where
\begin{eqnarray}
\zeta_{t} = \left(\int_{s=t}^\infty \int_{|x|>0}\left(\phi_s(x)  +\int_{s_1 = 0}^{t}\int_{|x_1|>0}\phi_{s\,s_1}(x,x_1) \, \tilde{N}( {\rm d}x_1, {\rm d}s_1) \right)\, \tilde{N}( {\rm d}x, {\rm d}s)\right)^2,
\end{eqnarray}
\begin{eqnarray}
\eta_{t} = \left(\int_{s=t}^\infty \int_{|x|>0}\int_{s_1 = t}^{s}\int_{|x_1|>0}\phi_{s\,s_1}(x,x_1) \, \tilde{N}( {\rm d}x_1, {\rm d}s_1) \, \tilde{N}( {\rm d}x, {\rm d}s)\right)^2,
\end{eqnarray}
\begin{align}
\theta_{t} &= \int_{s=t}^\infty \int_{|x|>0}\left(\phi_s(x)  +\int_{s_1 = 0}^{t}\int_{|x_1|>0}\phi_{s\,s_1}(x,x_1) \, \tilde{N}( {\rm d}x_1, {\rm d}s_1) \right)\, \tilde{N}( {\rm d}x, {\rm d}s) \nonumber \\
& \times
\int_{s=t}^\infty \int_{|x|>0}\int_{s_1 = t}^{s}\int_{|x_1|>0}\phi_{s\,s_1}(x,x_1) \, \tilde{N}( {\rm d}x_1, {\rm d}s_1) \, \tilde{N}( {\rm d}x, {\rm d}s)\, .
\end{align}
Then taking the conditional expectation of the above three terms and making use of the Ito isometry for Poisson random measure we have
\begin{eqnarray}
\mathbb{E}_t[\zeta_{t}] = \int_{s=t}^\infty \int_{|x|>0}\left(\phi_s(x)  +\int_{s_1 = 0}^{t}\int_{|x_1|>0}\phi_{s\,s_1}(x,x_1) \, \tilde{N}( {\rm d}x_1, {\rm d}s_1) \right)^2\, \nu( {\rm d}x) {\rm d}s,
\end{eqnarray}
\begin{eqnarray}
\mathbb{E}_t[\eta_{t}] = \int_{s=t}^\infty \int_{|x|>0}\mathbb{E}_t\left[\left(\int_{s_1 = t}^{s}\int_{|x_1|>0} \phi_{s\,s_1}(x,x_1) \, \tilde{N}( {\rm d}x_1, {\rm d}s_1)\right)^2 \right]\, \nu( {\rm d}x) {\rm d}s,
\label{eta}
\end{eqnarray}
\begin{align}
\mathbb{E}_t[\theta_{t}] &= \int_{s=t}^\infty \int_{|x|>0}\left(\phi_s(x)  +\int_{s_1 = 0}^{t}\int_{|x_1|>0}\phi_{s\,s_1}(x,x_1) \, \tilde{N}( {\rm d}x_1, {\rm d}s_1) \right)\, \nonumber \\
& \times
\mathbb{E}_t\left[\int_{s_1 = t}^{s}\int_{|x_1|>0}\phi_{s\,s_1}(x,x_1) \, \tilde{N}( {\rm d}x_1, {\rm d}s_1)\right] \, \nu( {\rm d}x) {\rm d}s\, .
\label{theta} 
\end{align}
We observe that \eqref{eta} can be simplified further by use of the Ito isometry and \eqref{theta} vanishes. Thus,  the pricing kernel takes the following form:
\begin{align}
\pi_t = &  \int_{s=t}^{\infty}\int_{x} \, \left( \phi_s(x) + \int_{s_1 = 0}^{t}\int_{|x_1|>0}\phi_{s\,s_1}(x,x_1) \, \tilde{N}( {\rm d}x_1, {\rm d}s_1) \right)^2 \nu( {\rm d}x)\,{\rm d}s  \nonumber \\ &
+ \, \int_{s=t}^{\infty}\int_{x} \,   \int_{s_1=t}^{s}\int_{x_1}\phi^{\,2}_{s\,s_1}(x,x_1) \, \nu( {\rm d}x_1)\,{\rm d}s_1 \,\nu( {\rm d}x)\,{\rm d}s \,.
\label{Pricing Kernel Expansion}
\end{align}
This  formula allows one to work out expressions for the discount bond prices, the short rate, and the market price of risk. Now, the price at time $t$ of a bond with maturity $T$  is given by the standard valuation formula \eqref{Constantinides formula}.
A calculation making use of \eqref{LI Isometry} then shows that
\begin{align}
 \mathbb{E}_t\,[\pi_T]  = &  \int_{s=T}^{\infty}\int_{x} \, \left( \phi_s(x) + \int_{s_1 = 0}^{t}\int_{|x_1|>0}\phi_{s\,s_1}(x,x_1) \, \tilde{N}( {\rm d}x_1, {\rm d}s_1) \right)^2 \nu( {\rm d}x)\,{\rm d}s  \nonumber \\ &
+ \, \int_{s=T}^{\infty}\int_{x} \,   \int_{s_1=t}^{s}\int_{x_1}\phi^{\,2}_{s\,s_1}(x,x_1) \, \nu( {\rm d}x_1)\,{\rm d}s_1 \,\nu( {\rm d}x)\,{\rm d}s \, .
\label{Conditional pricing kernel}
\end{align}
Then by inserting \eqref{Pricing Kernel Expansion} and \eqref{Conditional pricing kernel} into \eqref{Constantinides formula},  we are able to determine the bond price explicitly in the general second-order chaos model. Similar results can be obtained for higher order levels in the chaos expansion and Brownian terms can be added in as well.
%
\section{Factorizable Second-Order Chaos Models}
\label{sec:CM2}
\noindent

\noindent As a special case of the second-order chaos model one can consider what we call factorizable models, corresponding to the situation where the deterministic second-order chaos coefficient factorizes into a product of the form 
\begin{eqnarray}
\phi_{s\,s_1}(x,x_1) = \beta_s(x) \,  \gamma_{s_1}(x_1) \, .
\end{eqnarray}
Under this simplifying assumption, the pricing kernel takes the form
\begin{align}
\pi_t &= \int_{s=t}^{\infty}\int_{x}\phi^2_s(x) \,\nu( {\rm d}x)\,{\rm d}s  +  \int_{s=t}^{\infty}\int_{x} \beta^2_s(x) \,\nu( {\rm d}x) \int_{s_1=t}^{s}\int_{x_1} \gamma^2_{s_1}(x_1)\,\nu( {\rm d}x_1)\,{\rm d}s_1 \,{\rm d}s \nonumber \\
& + 2 \int_{s=t}^{\infty}\int_{x} \phi_s(x) \int_{s_1=0}^{t}\int_{|x_1|>0} \beta_s(x) \, \gamma_{s_1}(x_1) \, \tilde{N}( {\rm d}x_1, {\rm d}s_1) \,\nu( {\rm d}x)\,{\rm d}s \nonumber \\
& +  \int_{s=t}^{\infty}\int_{x} \left(\int_{s_1=0}^{t}\int_{|x_1| > 0} \beta_s(x)\, \gamma_{s_1}(x_1)\, \tilde{N}( {\rm d}x_1, {\rm d}s_1)  \right)^2 \,\nu( {\rm d}x)\,{\rm d}s.
\end{align}
If we split the integration with respect to $s_1$ in the first term this gives
\begin{align}
\pi_t &= \int_{s=t}^{\infty}\int_{x}\phi^2_s(x) \,\nu( {\rm d}x)\,{\rm d}s \nonumber \\
&+\int_{s=t}^{\infty}\int_{x} \beta^2_s(x) \,\nu( {\rm d}x) \left(\int_{s_1=0}^{s}\int_{x_1} \gamma^2_{s_1}(x_1)\,\nu( {\rm d}x_1) - \int_{s_1=0}^{t}\int_{x_1} \gamma^2_{s_1}(x_1)\,\nu( {\rm d}x_1)\,{\rm d}s_1 \right) \,{\rm d}s  \nonumber \\
& + 2 \int_{s=t}^{\infty}\int_{x} \phi_s(x)\,\beta_s(x) \,\nu( {\rm d}x)\,{\rm d}s \int_{s_1=0}^{t}\int_{|x_1|>0}  \gamma_{s_1}(x_1) \, \tilde{N}( {\rm d}x_1, {\rm d}s_1) \nonumber \\
& +  \int_{s=t}^{\infty}\int_{x} \left(\int_{s_1=0}^{t}\int_{|x_1| > 0} \beta_s(x)\, \gamma_{s_1}(x_1)\, \tilde{N}( {\rm d}x_1, {\rm d}s_1)  \right)^2 \,\nu( {\rm d}x)\,{\rm d}s.
\end{align}
Then we find that the pricing kernel is linear combination of a pair of martingales.  More precisely, if we define the process $\{ M_t \}_{t \geq 0}$ by setting
\begin{eqnarray}
M_t = \int_{s_1 = 0}^{t}\int_{|x_1|>0}\gamma_{s_1}(x_1) \, \tilde{N}( {\rm d}x_1, {\rm d}s_1) \, ,
\end{eqnarray}
we find that $\{M_t\}$ is a square-integrable martingale for which the associated predictable quadratic variation process $\{ Q_t \}_{t \geq 0}$ is given by
\begin{eqnarray}
Q_t = \int_{s_1 = 0}^{t}\int_{x_1}\gamma^{\,2}_{s_1}(x_1) \, \nu( {\rm d}x_1) \, {\rm d}s_1 \, .
\end{eqnarray}
Then one can check that the process $\{ M^{\,2}_t - Q_t \}_{t \geq 0}$ is also a martingale, and that the pricing kernel takes the form
\begin{eqnarray}
\pi_t = A_t + B_t \,M_t + C_t\,(M^2_t - Q_t) \, ,
\label{Factorizable pricing kernel}
\end{eqnarray}
where the deterministic coefficients $A_t$, $B_t$ and $C_t$ are defined as follows\,:
\begin{eqnarray}
A_t = \int_{t}^{\infty}\int_{x}\phi^{\,2}_s(x)\, \nu( {\rm d}x) \, {\rm d}s + \int_{t}^{\infty}\int_{x} \beta^{\,2}_s(x)   \, \nu( {\rm d}x)  \int_{s_1 = 0}^{s}\int_{x_1} \gamma^{\,2}_{s_1}(x_1)\, \nu( {\rm d}x_1) \, {\rm d}s_1  \, {\rm d}s \, , \nonumber
\end{eqnarray}
\begin{eqnarray}
B_t = 2\int_{t}^{\infty}\int_{x} \phi_s(x) \, \beta_s(x)\, \nu( {\rm d}x)\,{\rm d}s \,, \quad 
C_t = \int_{t}^{\infty}\int_{x}\beta^{\,2}_s(x) \, \nu( {\rm d}x)\,{\rm d}s \, .
\end{eqnarray}
Taking the conditional expectation of $\pi_T$, and using the martingale condition, we obtain
\begin{eqnarray}
\mathbb{E}_t[\pi_T] = {A_T} + B_T \,M_t + C_T\,(M^2_t - Q_t) \, .
\label{Conditional Factorizable pricing kernel}
\end{eqnarray}
Equations  \eqref{Factorizable pricing kernel} and \eqref{Conditional Factorizable pricing kernel} then show that the bond price is given by a rational function of $M_t$. More specifically, we see that $P_{tT}$ takes the form of a ratio of a pair of quadratic polynomials in $M_t$ with deterministic coefficients\,:
\begin{eqnarray}
P_{tT} = \mathds{1}\{t<T\} \, \frac { {A_T} + B_T \,M_t + C_T\,(M^2_t - Q_t)} {A_t + B_t \,M_t + C_t\,(M^2_t - Q_t)} \, .
\label{bond price as ratio}
\end{eqnarray}
Alternatively, one can view the bond price as being given by a linear rational function of a pair of martingales. Remarkably, the structure of the bond price system is identical to that arising in the factorizable second-order Brownian chaos model \cite{Rafailidis, Hughston Rafailidis 2005, Brody Hughston 2004}, which also exhibits a linear rational structure. 

We proceed to consider the calibration of the factorizable second-order chaos model to market data. The first requirement that one can impose on any interest rate model with freely specifiable time-dependent degrees of freedom is that we should be able to calibrate the model to an arbitrarily specified initial yield curve. Thus, in the present context we assume that the initial discount function $\{\bar P_{0t}\}_{t \geq 0}$ is given in the form of a strictly decreasing function admitting a continuous first derivative. The problem is to choose the deterministic functions $\{\phi_t(x)\}$, $\{\beta_t(x)\}$, $\{\gamma_t(x)\}$ in such a way that for $t \geq 0$ we have
\begin{eqnarray}
\bar P_{0t} = A_t  \, /  A_0 \, .
\label{Initial term structure}
\end{eqnarray}
First, we notice that we can rescale $\{\phi_t(x)\}$ and $\{\beta_t(x)\}$ by a common constant factor, without changing the resulting bond prices, in such a way that $A_0 = 1$. Once this is done, we must choose the renormalized functions  $\{\phi_t(x)\}$, $\{\beta_t(x)\}$, $\{\gamma_t(x)\}$ so that
\begin{eqnarray}
\bar P_{0t} = \int_{t}^{\infty} \int_{x} \phi^{\,2}_s(x) \, \nu({\rm d}x) \, {\rm d}s + \int_{t}^{\infty} \left[ \int_{x} \beta^{\,2}_s(x)  \nu({\rm d}x) \int_{s_1 = 0}^{s} \int_{x_1} \gamma^{\,2}_{s_1}(x_1)  \nu({\rm d}x_1) \,{\rm d}s_1 \right] {\rm d}s.
\label{Initial term structure formula}
\end{eqnarray}
The next step is to differentiate each side of this equation with respect to $t$ and define the instantaneous forward rate
\begin{eqnarray}
f_{0t} = \,- \frac{{\rm d} \log \bar P_{0t}}{{\rm d} t}\,.
\end{eqnarray}
Then the calibration condition takes the form 
\begin{eqnarray}
f_{0t} \, \bar P_{0t} =  \int_{x} \phi^{\,2}_t(x) \, \nu({\rm d}x)  +  \int_{x} \beta^{\,2}_t(x) \, \nu({\rm d}x) \, \int_{0}^{t} \int_{x_1} \gamma^{\,2}_{s_1}(x_1) \, \nu({\rm d}x_1) \,{\rm d}s_1\,.
\end{eqnarray}
Let us  regard the function $\{\gamma_t(x)\}$ as being a freely specifiable functional degree of freedom of the model satisfying 
$\lim_{t \to \infty} h_t  < \infty$ and $h_t > 0$ for $t>0$, where
\begin{eqnarray}
h_t =  \int_{0}^{t} \int_{x_1} \gamma^{\,2}_{s_1}(x_1) \, \nu({\rm d}x_1) \,{\rm d}s_1\,.
\label{h_t}
\end{eqnarray}
Then we set $\theta^2_t(x) = \beta^2_t(x)\,h_t$. The problem is thus to find $\{\phi_t(x)\}$ and $\{\theta_t(x)\}$ such that 
\begin{eqnarray}
f_{0t} \, \bar P_{0t} =  \int_{x} \left[ \phi^{\,2}_t(x) + \theta^{\,2}_t(x) \right] \, \nu({\rm d}x)\,
\label{f_t}
\end{eqnarray}
for all $t \geq 0$. This equation can be solved by setting
\begin{eqnarray}
\phi^{\,2}_t(x) = f_{0t}\, \bar P_{0t} \, p_t(x) \,, \quad \theta^{\,2}_t(x) = f_{0t}\, \bar P_{0t} \, q_t(x)\,,
\end{eqnarray}
where the functions $\{p_t(x)\}$ and $\{q_t(x)\}$ are taken to be non-negative and such that 
\begin{eqnarray}
\int_{x} \left[\, p_t(x) + q_t(x) \,\right]  \nu({\rm d}x) = 1, 
\label{condition on p_t and q_t}
\end{eqnarray}
for all $t\geq 0$. The existence of functions satisfying \eqref{condition on p_t and q_t} can be demonstrated by 
\begin{eqnarray}
p_t(x) = \frac{p\,\,(1\wedge x^2)}{\int_{x}(1\wedge x^2) \, \nu({\rm d}x)} \, , \quad q_t(x) = \frac{q\,\,(1\wedge x^2)}{\int_{x}(1\wedge x^2) \, \nu({\rm d}x)} \, ,
\end{eqnarray}
where $p$ and $q$ are positive constants such that $p+q=1$; but clearly one can also find more general functions satisfying \eqref{condition on p_t and q_t}. With these conditions imposed we have satisfied \eqref{f_t}. In summary, therefore, we have established the following: 
\vspace{2mm}
\begin{Proposition} \, In a factorizable second-order chaos model, let the initial instantaneous forward rate curve be given as a non-negative continuous function $\{f_{0t}\}$.  Then a solution for the calibration of the model to this initial data is obtained by letting $\{\gamma_t(x)\}$ be given freely, defining $\{h_t\}$ as in 
\eqref{h_t}, and letting  $\{\phi_t(x)\}$ and $\{\beta_t(x)\}$ be  given by
\begin{eqnarray}
\phi^2_t(x) = f_{0t}\, \bar P_{0t}\, p_t(x) \, , \quad \beta^2_t(x) = \frac{1}{h_t} f_{0t}\, \bar P_{0t}\, q_t(x) \, ,
\end{eqnarray}
where $\{p_t(x)\}$ and $\{q_t(x)\}$ are non-negative and satisfy   \eqref{condition on p_t and q_t}\,. 
\end{Proposition}
\vspace{2mm}
\noindent The remaining functional degrees of freedom can be then used to calibrate the model to the prices of other market instruments by methods similar to those employed in \cite{Tsujimoto 2010, Grasselli Tsujimoto 2011} in the Brownian case. One can also use the L\'evy measure itself as a functional degree of freedom for the purpose of calibration, as discussed for example in \cite{Bouzianis Hughston 2019}.

\section{L\'evy-Ito models for Foreign Exchange}
\label{sec:FEM}
\noindent
We consider a system of exchange rates $\{F^{ij}_t\}_{t \geq 0}$ for $N$ currencies $(i,j =1,...,N)$. Here $F^{ij}_t$ denotes the price at time $t$ of one unit of currency $i$ expressed in units of currency $j$.  
As in our earlier considerations, we let $N({\rm d}x, {\rm d}t)$ denote the Poisson random measure associated with an underlying $n$-dimensional L\'evy process with L\'evy measure  $\nu({\rm d}x)$. Typically, we require that $n \geq N-1$ in order to ensure that the model has sufficient freedom. The idea is that we fix one of the currencies as a base currency (or ``domestic'' currency) and we consider the dynamics of the prices of the $N-1$ remaining currencies when these prices are expressed in units of the base currency. Therefore, we would like the state space of the L\'evy-Ito process to be at least of dimension $N-1$. For instance, in the case of three currencies, an underlying two-dimensional L\'evy process is the necessary minimal structure. 

To construct the general form of the exchange rate matrix we model a system of $N$ pricing kernels $\{\pi^i_t\}_{t \geq 0}$, one for each currency, by setting
\begin{align}
\pi^i_t = \pi^i_0 \exp \left[-\int_0^t r^i_s \, {\rm d}s  - \int_{0}^{t} \int_{|x|>0} \lambda^i_s (x) \, \tilde{N}({\rm d}x, {\rm d}s) - \int_{0}^{t} \int_{x} \left(\re^{-\lambda^i_s(x)} -1 + \lambda^i_s(x) \right) \, \nu({\rm d}x) \, {\rm d}s  \right]. \nonumber
\\
\label{Levy Ito FX pricing kernel}
\end{align} 
Here we have suppressed the $n$-dimensional Brownian component of the L\'evy-Ito process; the general case including the Brownian terms can be easily reconstructed.  The same conditions are imposed on each of the pricing kernels here were imposed on the pricing kernel in Section III, except that here it will be convenient to give each pricing kernel a distinct initial value. The idea of having a pricing kernel for each currency is that the pricing kernel for a given currency can be used to value assets that are priced in that currency. If two economies based on separate currencies are economically independent, then we would expect the corresponding pricing kernels to be independent in the probabilistic sense. In reality, of course, the major economies are interdependent in various complex ways, so in general we expect the pricing kernels associated with various currencies to exhibit correlations.  Then the fundamental property of the exchange rate matrix is that for each currency pair the relevant component of the matrix is given by the ratio of the pricing kernels associated with the two currencies \cite{Flesaker Hughston 1997, Rogers 1997, Lipton 2001}. More precisely, we have
\begin{eqnarray} 
F^{ij}_t = \pi^i_t \, / \pi^j_t \, .
\label{Fundamental FX property}
\end{eqnarray}
We are then led to the following:
\vspace{2mm}
\begin{Proposition} \, In a general L\'evy-Ito setting, the exchange rate matrix takes the form
\begin{align}
F^{ij}_t = F^{ij}_0 &\exp \left[\int_0^t (r^{j}_s - r^{i}_s  + R^{ij}_s) \, {\rm d}s + \int_{0}^{t} \int_{|x|>0} \sigma^{ij}_s (x) \, \tilde{N}({\rm d}x, {\rm d}s)\right] \nonumber \\
& \times \exp \left[ -\int_{0}^{t} \int_{x} \left(\re^{\sigma^{ij}_s (x)} -1 - \sigma^{ij}_s (x) \right) \, \nu({\rm d}x) \, {\rm d}s  \right],
\label{FX price}
\end{align} 
with initial exchange rate $F^{ij}_0 = \pi^i_0 \, / \pi^j_0$\,, where the excess rate of return is given by
\begin{eqnarray} 
R^{ij}_t = \int_{x}\left(\re^{\sigma^{ij}_t(x) }-1\right)\left(1-\re^{-\lambda^{j}_t(x)}\right) \,\nu({\rm d}x)\,,
\label{FX Excess Rate}
\end{eqnarray}
and for the exchange rate volatility one has
\begin{eqnarray} 
\sigma^{ij}_t(x) = \lambda^{j}_t(x) - \lambda^{i}_t(x) \, .
\label{FX volatility}
\end{eqnarray} 
\end{Proposition}
\vspace{2mm}
\noindent \textit{Proof}. If we combine (\ref{Levy Ito FX pricing kernel}) and (\ref{Fundamental FX property}), a straightforward  calculation gives 
\begin{align}
F^{ij}_t = F^{ij}_0 &\exp \left[\int_0^t (r^{j}_s - r^{i}_s ) \, {\rm d}s + \int_{0}^{t} \int_{|x|>0} \sigma^{ij}_s (x) \, \tilde{N}({\rm d}x, {\rm d}s)\right] \nonumber \\
& \times \exp \left[ -\int_{0}^{t} \int_{x} \left(\re^{-\lambda^i_s(x)} - \re^{-\lambda^j_s(x)}+ \lambda^i_s(x) - \lambda^j_s(x)\right) \, \nu({\rm d}x) \, {\rm d}s  \right].
\label{tentative form for FX}
\end{align} 
Next we observe that by \eqref{lambda condition} and \eqref{Lambda sup sup 3} it holds that
\begin{eqnarray} 
\mathbb{P} \left[ \int_{0}^{t} \int_{|x|\geq 1}  \re^{-\lambda^i_s (x)}\, \nu({\rm d}x) \, {\rm d}s < \infty \right] = 1 \,, \quad 
\mathbb{P} \left[ \sup_{0 \leq s \leq t} \, \sup_{0 \leq |x| < \infty }  \lambda^i_s(x) < \infty \right] = 1\,
\end{eqnarray}
for $i = 1, \dots, n$, from which we deduce that 
\begin{eqnarray} 
\mathbb{P} \left[ \int_{0}^{t} \int_{|x|\geq 1}  \re^{\lambda^j_s (x)-\lambda^i_s (x)}\, \nu({\rm d}x) \, {\rm d}s < \infty \right] = 1 \,,
\end{eqnarray}
for $i,j = 1, \dots, n$. It follows that $R^{ij}_t < \infty$ almost surely for $i,j = 1, \dots, n$ and for all $t\geq 0$, and hence that we can regroup the terms in \eqref{tentative form for FX} to obtain \eqref{FX price}. \hfill $\square$

\vspace{.2cm}
In particular, one sees from 
Proposition  7 that once the short rates and the risk aversion processes have been specified for each of the currencies, along with the initial exchange rates, then the exchange rate dynamics are completely determined. It is interesting to observe that for each pair of currencies the exchange rate volatility decomposes into a pair of terms, one for each of the two currencies. The significance of this fact is that it seems that one cannot model exchange rate volatility ``directly'' by simply positing an {\em ad hoc} form for $\{\sigma^{ij}_t(x)\}$. 
There is, of course, a substantial literature devoted to attempts at modelling exchange rate volatility, and it has to be said that much of this is carried out without taking into account the risk aversion functions associated with each currency and the decomposition given by equation (\ref{FX volatility}). We claim that such investigations are  more natural if the modelling is pursued at the level of the individual risk aversion functions for the various currencies. 

We turn now to consider the excess rate of return, which in a pure-jump L\'evy-Ito model for foreign exchange takes the form \eqref{FX Excess Rate}.
It is interesting to ask if it is possible for $R^{ij}_t$ to be positive for all currency pairs. If a model has this property, we say that it satisfies the Siegel condition. Siegel \cite{Siegel} seems to have been the first to identify the seemingly paradoxical fact that in a stochastic model it is consistent, for example, for  the {\textsc EUR-USD} exchange rate and the USD-EUR exchange rate to exhibit positive excess rates of return simultaneously, even though the exchange rates are inverses of one another. The problem of determining whether it is possible for $R^{ij}_t$ to be positive for all currency pairs in a setting with $N$ currencies involves showing that $N\,(N-1)$ different exchange rates have positive excess rates of return. The intuition is that if any of these rates were to show a negative excess rate of return, then investors would sell off the overpriced currency until a new price level was reached with the property that the excess rate of return was no longer negative.

We shall prove the existence of $N$-currency models of the L\'evy type in which all $N\,(N-1)$ excess rates of return are strictly positive. The argument is non-trivial even in the Brownian case, so we consider that first. Then we look at a model involving $n$ identical copies of a given L\'evy process (this example is based on a suggestion made by an anonymous reviewer). Next we construct a class of $N$-currency Merton-type models, i.e.~compound Poisson with Gaussian jumps. Finally, we consider an $N$-currency model driven by an $n$-dimensional generalization of the variance gamma process. On the basis of these examples we are led to conjecture that the Siegel condition can be satisfied in a broad class of L\'evy-Ito models. 

\vspace{0.3cm}

\noindent \textit{Geometric Brownian motion model}. In the Brownian case we let $\{F^{ij}_t\}$ denote a set of exchange rates between $N$ currencies $(i=1,\,\dots \,,\,N)$ driven by a family of $n$ independent Brownian motions. The pricing kernel for currency $i$ is taken to be of the form  
\begin{equation}
\pi^i_t=\pi^i_0\,\exp \left[-r^i\,t-{\lambda}^i\cdot{W}_t-\tfrac{1}{2}\,{\lambda}^i\cdot{\lambda}^i\,t\right] ,
\end{equation}
where $r^i$ is the interest rate for currency $i$, ${\lambda}^i$ is a vector in $\mathbb{R}^{n}$ for each value of $i$, and $\{{W}_t\}$ is a Brownian motion taking values in $\mathbb{R}^{n}$. The dot denotes the usual inner product between vectors in $\mathbb{R}^{n}$. It follows from \eqref{Fundamental FX property} that 
\begin{equation}
F^{ij}_t=F^{ij}_0\,\exp \left[(r^j-r^i)\,t+R^{ij}\,t+{\sigma}^{ij}\cdot{W}_t-\tfrac{1}{2}\,{\sigma}^{ij}\cdot{\sigma}^{ij}\,t\right] ,
\end{equation}
where ${\sigma}^{ij}={\lambda}^j-{\lambda}^i$ and $R^{ij}={\sigma}^{ij}\cdot{\lambda}^j$. Thus, the question is whether we can choose the ${\lambda}^i$ vectors $(i=1,\,\dots \,,\,N)$ in such a way that for all $i,\,j$ ($i\neq j$) one has
\begin{equation}\label{Positive ERR BC}
\left({\lambda}^j-{\lambda}^i\right) \cdot {\lambda}^j>0\,.
\end{equation}
The answer turns out to be yes, as the following construction shows. 
Let ${\lambda}^i$ $(i=1,\,\dots \,,\,N)$ be a set of $N$ distinct vectors, each of the same length, so we have ${\lambda}^i \cdot {\lambda}^i= L^2$ for some fixed length $L>0$, for all $i$. Then for each pair $i,\,j$ ($i\neq j$) we have 
\begin{equation}
{\lambda}^i\cdot{\lambda}^j=L^2\,\cos \theta^{ij}, 
\end{equation}
where $\theta^{ij}$ is the angle between the two vectors. We have assumed that the $N$ equal-length vectors are distinct, so $\theta^{ij}\neq0$ for each pair $i,\,j$ ($i\neq j$). 
As a consequence we see that $\cos \theta^{ij} <1$ for each such pair, and this leads to the result \eqref{Positive ERR BC}. 
Thus we have demonstrated the existence of $N$-currency geometric Brownian motion models for which the Siegel condition holds for each currency pair. Putting the matter somewhat more generally, we have shown that for any $N \geq 2$ one can construct non-trivial arbitrage-free foreign exchange models with the property that the excess rate of return is positive for each of the $N(N-1)$ exchange rates. 
\vspace{0.3cm}

\noindent \textit{Independent identical copies of a L\'evy process}. As another example, let $\{X^i_t\}$, $i = 1, \dots, N$, be a collection of $N$ independent identical copies of a L\'evy process $\{X_t\}$. We assume that 
$\{X_t\}$ admits exponential moments and hence that we can form the L\'evy exponent
\begin{equation}
\psi(\alpha) =  \frac {1}{t} \log \mathbb E \left[ \exp (\alpha X_t) \right] 
\end{equation}
for $\alpha$ in some nontrivial interval containing the origin. Then we form a set of $N$ independent pricing kernels $\{\pi^i_t\}$ of the form
\begin{equation}
\pi^i_t=\pi^i_0\,\exp \left[-r^i\,t - {\lambda}X_t^i - \psi(-\lambda)\, t \right] . 
\end{equation}
Hence, for the associated exchange rate system we have
\begin{equation}
F^{ij}_t=F^{ij}_0\,\exp \left[(r^j-r^i)\,t+R^{ij}\,t+{\lambda}(X_t^j - X_t^i)-(\psi(\lambda) + \psi(-\lambda) ) \, t\right] ,
\end{equation}
where $R^{ij} = \psi(\lambda) + \psi(-\lambda)$ for all $i,\,j$ ($i\neq j$). Thus all of the excess rates of return are the same. Moreover, since $X_t^j - X_t^i$ has mean zero for each currency pair, it follows by Jensen's inequality that the excess rates of return are positive. Thus we have an $N$-currency model driven by $N$ L\'evy processes with the property that the excess rate of return is positive for each currency pair. This example is a bit odd in the sense that all of the market prices of risk are the same and all of the exchange rate volatilities are the same. Nevertheless, it demonstrates the principle that the Siegel condition can hold consistently across all currency pairs in an multi-currency L\'evy market. If the L\'evy process here happens to be a Brownian motion, then the present example reduces to a special case of the previous example. 

\vspace{0.3cm}

\noindent \textit{Merton model}. We proceed to consider an $N$-currency model driven by an $(N-1)$-dimensional pure-jump process of the Merton type \cite{Merton 1976}. It will suffice to show the details of a three-currency model driven by a two-dimensional Merton process; the reader will be able to supply the straightforward generalization to the $N$-currency situation. 
Thus, we consider a two-dimensional compound Poisson process given by a pair of processes of the form
\begin{equation}
\xi^1_t=\sum_{\kappa=1}^{N_t}X_{\kappa}    \,,\quad \xi^2_t=\sum_{\kappa=1}^{N_t}Y_{\kappa}\,,
\end{equation}
where the $(X_{\kappa})_{\kappa \in \mathbb N}$ constitute an independency of identically distributed random variables, the $(Y_{\kappa})_{\kappa \in \mathbb N}$ constitute another such independency, and $\{N_t\}_{t\geq 0}$ is an independent Poisson process. For fixed $\kappa$, the random variables $X_{\kappa}$ and $Y_{\kappa}$ are not necessarily independent, and for a typical such pair 
$X,Y$ we write 
\begin{equation}
\phi(\alpha,\beta)=\mathbb{E}\left[e^{\alpha\,X+\beta\,Y}\right] ,
\end{equation}
under the assumption that the moment generating function is finite for a non-trivial range of values of 
$\alpha$ and $\beta$. The associated bivariate L\'evy exponent is then defined by
\begin{equation}
\psi(\alpha,\beta)=\frac{1}{t}\,\log\mathbb{E}\left[e^{\alpha\,\xi^1_t+\beta\,\xi^2_t}\right] ,
\end{equation}
and a calculation shows that 
\begin{equation}
\psi(\alpha,\beta)=m\,\left[\phi(\alpha,\beta)-1\right] ,
\end{equation}
where $m$ is the intensity of the underlying Poisson process. Thus, in this example the jump times of the two processes coincide, but the jump sizes are random and generally distinct. 
In the case of a Merton-type model, we have $X,Y\sim N(\mu_1,\mu_2,\sigma_1,\sigma_2,\rho)$, and hence 
\begin{equation}
\psi(\alpha,\beta)=m\,\left[\exp\left(\alpha\,\mu_1+\beta\,\mu_2+\tfrac{1}{2}\,\alpha^2\,\sigma_1^2+\tfrac{1}{2}\,\beta^2\,\sigma_2^2+\alpha\,\beta\,\sigma_1\,\sigma_2\,\rho\right) - 1 \right] .
\label{Merton psi}
\end{equation}
We introduce the vectors 
\begin{equation}
{\xi}_t=\left(\xi^1_t,\,\xi^2_t\right)
\label{xi vector}
\end{equation}
and 
\begin{equation}
{\lambda}^1=\left(a_1,\,b_1\right),\quad {\lambda}^2=\left(a_2,\,b_2\right),\quad{\lambda}^3=\left(a_3,\,b_3\right).
\label{lambda vectors}
\end{equation}
For the pricing kernels associated with the three currencies we set
\begin{equation}\label{Pricing Kernel for FX PJ}
\pi^i_t=\pi^i_0\,\exp\left[-r^i\,t-{\lambda}^i\cdot{\xi}_t-\psi\left({\lambda}^i\right)\,t\right] ,
\end{equation}
for $i=1,\,2,\,3$.
The exchange rate matrix is then given by
\begin{equation}
F^{ij}_t=F^{ij}_0\,\exp\left[\left(r^j-r^i\right)t +R^{ij}\,t+\left({\lambda}^j-{\lambda}^i\right)\cdot{\xi}_t-\psi\left({\lambda}^j-{\lambda}^i\right)\,t\right] ,
\label{exchange rate matrix}
\end{equation}
where 
\begin{equation}
R^{ij}=\psi\left({\lambda}^j-{\lambda}^i\right)+\psi\left(-{\lambda}^j\right)-\psi\left(-{\lambda}^i\right).
\label{excess RoR}
\end{equation}
It follows by \eqref{Merton psi} and \eqref{excess RoR} that to establish the existence of a three-currency pure-jump model satisfying the Siegel condition it suffices to show that one can choose the parameters of the bivariate normal distribution along with the three vectors 
$\{{\lambda}^i\}_{i=1,2,3}$ so that 
\begin{equation}\label{Condition two-JD Ex}
\re^{\left({\lambda}^j-{\lambda}^i\right)\,{\mu}^\text{T}\,+\,\frac{1}{2}\,\left({\lambda}^j-{\lambda}^i\right)\,{C}\,\left({\lambda}^j-{\lambda}^i\right)^\text{T}}+\re^{-{\lambda}^j\,{\mu}^\text{T}\,+\,\frac{1}{2}\,\left({\lambda}^j\right)\,{C}\,\left({\lambda}^j\right)^\text{T}}>\re^{-{\lambda}^i\,{\mu}^\text{T}\,+\,\frac{1}{2}\,\left({\lambda}^i\right)\,{C}\,\left({\lambda}^i\right)^\text{T}}+1\,,
\end{equation}
where ${\mu}=(\mu_1,\,\mu_2)$, $(\,\cdot\,)^\text{T}$ denotes the transpose operation, and ${C}$ is the covariance matrix of the $N (\mu_1,\mu_2,\sigma_1,\sigma_2,\rho)$ distribution.
To construct an explicit example, let us assume that $\mu_1=0$, $\mu_2=0$, $\sigma_1 = 1$, $\sigma_2 = 1$, and $\rho=0$. Then condition \eqref{Condition two-JD Ex} takes the form
\begin{equation}
\re^{\frac{1}{2}\,\left[(a_j\,-\,a_i)^2\,+\,(b_j\,-\,b_i)^2\right]}+\re^{\frac{1}{2}\,\left(a_j^2\,+\,b_j^2\right)}>\re^{\frac{1}{2}\,\left(a_i^2\,+\,b_i^2\right)}+1\,,
\label{Condition two-JD Ex in detail}
\end{equation}
where $a_i = (a_1, a_2, a_3)$ and $b_i = (b_1, b_2, b_3)$. The inequality \eqref{Condition two-JD Ex in detail} is manifestly satisfied if we choose the vectors $\{{\lambda}^i\}_{i=1,2,3}$ so that they are distinct and of equal length; that is to say, 
\begin{equation}
{{\lambda}^1} \neq {{\lambda}^2}, \quad {{\lambda}^1} \neq {{\lambda}^3}, \quad
{{\lambda}^2} \neq {{\lambda}^3}, 
\end{equation}
and
\begin{equation}
\norm{{\lambda}^1}=\norm{{\lambda}^2}=\norm{{\lambda}^3}. 
\end{equation}
For then we have 
\begin{equation}
a_i^2+b_i^2 = a_j^2+b_j^2
\end{equation}
for each currency pair,  but also
\begin{equation}
(a_j-a_i)^2+(b_j-b_i)^2 > 0 \, ,
\end{equation}
and hence \eqref{Condition two-JD Ex in detail}. Thus we have demonstrated the existence of a non-trivial three-currency finite-activity pure-jump L\'evy model satisfying the Siegel condition for all six exchange rates. The corresponding construction for any number of currencies is similar. 

\vspace{0.3cm}

\noindent \textit{Variance-gamma model}. An interesting example of an infinite activity L\'evy process leading to a foreign exchange model satisfying the Siegel condition for any number of currencies can be obtained as follows. We present the three-currency case in full. 
First, let us recall a few details of the theory of the variance-gamma process \cite{Madan Seneta 1990, Madan Milne 1991, Madan Carr Chang 1998}. Let $\{\Gamma_t\}_{t\geq 0}$ be a gamma process for which the parameters are chosen such that $\mathbb{E}\left[\Gamma_t\right]=t$, and $\text{Var}\left[\Gamma_t\right]=t/m$. We shall refer to such a process as a standard gamma subordinator with intensity $m$, following \cite{Brody Hughston Macrina 2008, Brody Hughston Mackie 2012}. For further aspects of the gamma process see \cite{Dicksen, Dufresne, Yor}.  Then by a variance-gamma process with intensity $m$, we mean a process $\{\xi_t\}_{t\geq 0}$ of the form $\xi_t=W_{\Gamma_t}$, where $\{W_t\}_{t\geq 0}$ is a standard Brownian motion and $\{\Gamma_t\}_{t\geq 0}$ is an independent standard gamma subordinator with intensity $m$. It is a straightforward exercise to check that 
\begin{equation}
\psi(\alpha)= -m\,\log\left(1-\frac{\alpha^2}{2\,m}\right),
\end{equation}
for $\alpha$ such that
\begin{equation}
-\sqrt{2\,m}<\alpha<\sqrt{2\,m}\,.
\end{equation}
In what follows we consider a three-currency exchange-rate system driven by a generalization of the variance-gamma process. Let $\{X_t\}_{t\geq 0}$, and $\{Y_t\}_{t\geq 0}$ be independent Brownian motions, let $\{\Gamma_t\}_{t\geq 0}$ be an independent standard gamma subordinator with intensity $m$, and set
\begin{equation}
\xi^1_t=X_{\Gamma_t}\,,\quad \xi^2_t=Y_{\Gamma_t}\,.  
\end{equation}
Then the vector  $\{\xi^1_t,\,\xi^2_t\}_{t\geq 0}$ is a two-dimensional L\'evy process, and  the associated L\'evy exponent is given by 
\begin{equation}
\psi(\alpha,\beta)= -m\,\log\left(1-\frac{\alpha^2+\beta^2}{2\,m}\right),
\end{equation}
for $\alpha, \beta$ such that
\begin{equation}
0 \leq \alpha^2 + \beta^2 <2m \,.
\label{norm condition}
\end{equation}
Let us define the vector ${\xi}_t$ as in equation \eqref{xi vector},  the vectors $\{{\lambda}^i\}_{i=1,2,3}$ as in equation \eqref{lambda vectors}, and  $\{\pi^i_t\}_{i=1,2,3}$ as in equation \eqref{Pricing Kernel for FX PJ}. Then the exchange rate matrix is given by \eqref{exchange rate matrix}, and the excess rate of return is given by \eqref{excess RoR}.  It should be evident by virtue of \eqref{norm condition} that in order for the pricing kernels to be well defined the risk aversion vectors must be such that 
\begin{equation}
\norm{{\lambda}^i}<\sqrt{2\,m}\, ,
\label{length condition}
\end{equation}
for $i = 1,2,3$. 
To construct a class of models satisfying the Siegel condition, we proceed thusly. Fix $m$, and let  the vectors $\{{\lambda}^i\}_{i=1,2,3}$ be distinct and of equal length. It follows immediately that for each currency pair we have
\begin{equation}
\psi\left(-{\lambda}^i\right) = \psi\left(-{\lambda}^j\right).
\end{equation}
Then the excess rate of return for each currency pair is well defined and strictly positive if and only if
\begin{equation}
\psi\left({\lambda}^j-{\lambda}^i\right) > 0 \, ,
\end{equation}
for all $i,j$ such that $i \neq j$, or equivalently
\begin{equation}
-m\,\log\left(1-\frac{\left( {\lambda}^j-{\lambda}^i \right)^2}{2\,m}\right)>0.
\end{equation}
Since the risk aversion vectors have been assumed to be distinct, it follows that $R^{ij} > 0$ for any currency pair if and only if
\begin{equation}
\norm{{\lambda}^i-{\lambda}^j}<\sqrt{2\,m} \, .
\end{equation}
Now, writing $L$ for the common length of the risk aversion vectors, we have
\begin{equation}
\left({\lambda}^i-{\lambda}^j\right)^2 = 2\,L^2\,(1-\cos \theta_{ij}),
\end{equation}
where $\theta_{ij}$ denotes the angle between ${\lambda}^i$ and ${\lambda}^j$. Hence, $R^{ij} > 0$ if and only if
\begin{equation}
\cos \theta_{ij} > 1 - \frac{m}{L^2} \, .
\end{equation}
On the other hand, since $L < \sqrt{2\,m}$ by \eqref{length condition}, a sufficient condition to ensure that the excess rate of return is positive for each currency pair  is 
\begin{equation}
\cos \theta_{ij} > \half \, , 
\end{equation}
that is to say, that the angle between each of the risk aversion vectors is less than sixty degrees. With this choice, we have thus shown the existence of a three-currency infinite activity L\'evy model satisfying the Siegel condition for all six exchange rates. In fact, if 

\begin{equation}
L^2 < \half \, m \, , 
\end{equation}
then the risk aversion vectors can be at any angle relative to each other and the Siegel condition will hold. The extension of the argument to four or more currencies is straightforward.


\begin{acknowledgements}
\noindent The authors thank D.~Brody, E.~Eberlein, M.~Grasselli, T.~Hurd, A.~Lokka, 
A.~Macrina, D.~Meier, B.~Meister, K.~Owari, G.~Peskir, M.~Pistorius, A.~Rafailidis, M.~Schweizer, and T.~Tsujimoto for helpful comments, along with participants at the 2019 Research in Options Conference, IMPA, Rio de Janeiro, and the University of Manchester financial mathematics seminar. We are grateful for support from (a) Timelineapp Limited, Basildon [GB], (b) the Fields Institute for Research in Mathematical Sciences,  the Aspen Center for Physics, and the Simons Foundation [LPH],  (c) the Natural Sciences and Engineering Research Council of Canada, grants RGPIN-2018-05705 and RGPAS-2018-522715 [SJ], and (d) Consejo Nacional de Ciencia y Tecnolog\'ia (CONACyT), Mexico, LMAX Exchange, London, and Oriel College, Oxford [LSB].
\\ \\ 
\end{acknowledgements}

\noindent {\bf References}

\end{document}